\documentclass[
    reprint,
    amsmath,
    amssymb,
    prstab,
    showkeys]{revtex4-2}

\usepackage[english]{babel}
\usepackage{hyperref}
\usepackage{mathtools}
\usepackage{siunitx}
\usepackage{float}
\usepackage{subcaption}
\usepackage{tikz}

\begin{document}

\title{Freeform shape optimization of a compact DC photo-electron gun\\ using
isogeometric analysis}

\author{Peter Förster}
\email{peter.foerster@tu-darmstadt.de}
\author{Sebastian Schöps}

\affiliation{Institute for Accelerator Science and Electromagnetic Fields,
Technische Universität Darmstadt, Schloßgartenstraße 8, 64289 Darmstadt,
Germany}

\author{Joachim Enders}
\author{Maximilian Herbert}

\affiliation{Institut für Kernphysik, Fachbereich Physik, Technische
Universität Darmstadt, Schloßgartenstraße 9, 64289 Darmstadt, Germany}

\author{Abele Simona}

\affiliation{Laboratory for Modeling and Scientific Computing, Politecnico
Milano, p.za Leonardo da Vinci 32, 20133 Milano, Italy}

\begin{abstract}
Compact DC high-voltage photo-electron guns are able to meet the sophisticated
demands of high-current applications such as energy recovery linacs. A main
design parameter for such sources is the electric field strength, which depends
on the electrode geometry and is limited by the field emission threshold of the
electrode material. In order to minimize the maximum field strength for optimal
gun operation, isogeometric analysis (IGA) can be used to exploit the
axisymmetric geometry and describe its cross section by non-uniform rational
B-splines, the control points of which are the parameters to be optimized. This
computationally efficient method is capable of describing CAD-generated
geometries using open source software (\textsc{GeoPDEs}, \textsc{NLopt},
\textsc{Octave}) and it can simplify the step from design to simulation. We
will present the mathematical formulation, the software workflow, and the
results of an IGA-based shape optimization for a planned high-voltage upgrade
of the DC photogun teststand \mbox{Photo-CATCH} at TU Darmstadt. The software
builds on a general framework for isogeometric analysis and allows for easy
adaptations to other geometries or quantities of interest. Simulations assuming
a bias voltage of \SI{-300}{\kilo\volt} yielded maximum field gradients of
\SI{9.06}{\mega\volt\per\meter} on the surface of an inverted insulator
electrode and below \SI{3}{\mega\volt\per\meter} on the surface of the
photocathode.
\end{abstract}

\maketitle

\section{Introduction} \label{sec:introduction}
Advanced applications of electron accelerators such as energy recovery linacs
(ERLs) \cite{sinclair2006, rao2006} require beams with high current and small
emittance, therefore placing sophisticated demands on electron sources.
State-of-the-art DC high-voltage photo-electron guns are promising candidates
for meeting these requirements \cite{maxson2014, sinclair2007}. The
electrostatic design for this type of source, in light of optimizing the beam
properties, has been discussed for many decades \cite{pierce1940, radley1958}.
For example, there exists extensive research focused on the optimization of
beam parameters depending on electron bunch parameters \cite{bazarov2005}. For
instance, the electrode geometry was optimized for beam emittance in
\cite{bazarov2011}, using a set of parameters that describe a few key geometric
features. In contrast, this paper is dedicated to optimizing the freeform shape
of the electrode in terms of CAD basis functions to minimize the electric field
strength, which has a crucial impact on field emission and thus still
represents a major design problem depending on the specific geometry of the
setup.

Low-level field emission can have a significant negative impact on the vacuum
conditions within the gun and may severely degrade beam quality and operational
lifetime \cite{grames2011}. High-level field emission can cause extensive
damage to both electrode and insulator, necessitating repair or even
replacement of the components. However, a high bias voltage is desired to
provide sufficient initial acceleration for the beam and to minimize the
emittance in spite of space charge effects. Since common negative bias voltages
of DC photo-electron guns are in the range of \SIrange{-100}{-500}{\kilo\volt}
\cite{breidenbach1994, poltoratska2011, garcia2019, nishimori2014},
the combination of such high voltages with a suitable electrode geometry and
material poses a great challenge for the design of compact guns. The decisive
limiting factor is the field emission threshold of the electrode material,
imposing a maximum electric field strength on the geometric design. While
increasing the curvature of the electrode surface reduces the field maximum,
the overall size of the electrode is limited since the surface area susceptible
to field emission should be kept small. Furthermore, a larger surface area also
downgrades the vacuum conditions \cite{stutzman2003} and lastly, a larger
electrode requires a larger vacuum chamber, which can be impractical due to
cost and space constraints. A promising approach is the so-called inverted
insulator geometry gun (IIGG) design \cite{breidenbach1994, adderley2010}, which
significantly reduces the size of the electrode by placing the high-voltage
insulator inside the vacuum chamber.

\begin{figure}
    \begin{center}
    \includegraphics[width=0.45\textwidth]{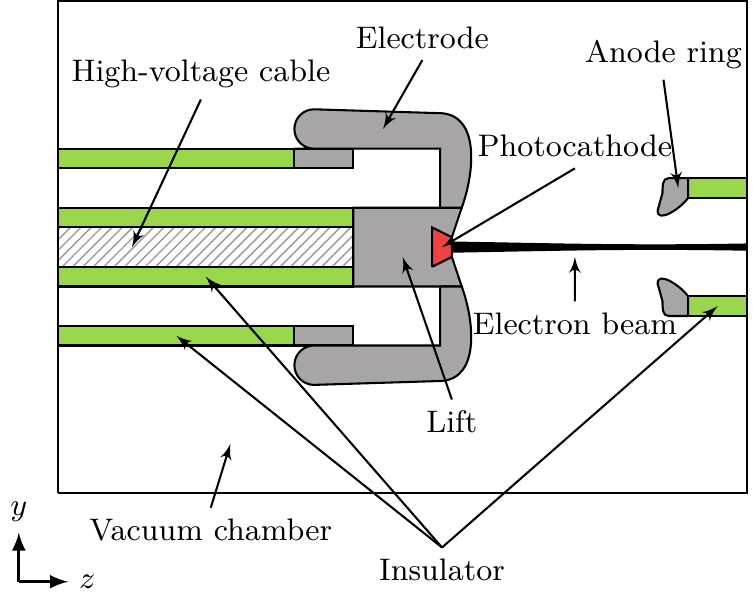}
    \caption{Basic components of the IIGG design in a longitudinal cross
             section of the vacuum chamber.}
    \label{fig:design}
    \end{center}
\end{figure}

Practical experience shows that unavoidable material impurities and limitations
in machining may cause significant variations in the field emission threshold.
It is therefore paramount to keep the maximum electric field strength of the
design well below the threshold. For stainless steel (1.4429 ESU), a commonly
used electrode material, the threshold estimate from operational observations
is \SI{10}{\mega\volt\per\meter} \cite{sinclair2001}. Other available
materials, such as niobium, titanium, and molybdenum possess a higher threshold
for field emission \cite{bastaninejad2012, furuta2005}, but are more expensive
and more difficult to machine. Common electrode designs range from simple
spherical and cylindrical forms to more complex geometries like the T-shaped
design used at JLab \cite{adderley2010}. At TU Darmstadt a test facility for
Photo-Cathode Activation, Test, and Cleaning using atomic-Hydrogen, \mbox{Photo-
CATCH}, which is also dedicated to DC photo-electron gun research and
development, has been established recently \cite{kurichiyanil2019}. It uses an
axisymmetric IIGG, featuring a two-part electrode consisting of a main
electrode body and an extendable lift for photocathode loading
\cite{herbert2018}. An upgrade from \SIrange{-60}{-300}{\kilo\volt} bias
voltage has been envisioned and is currently under development. In order to
meet design constraints concerning available space and chamber size, an
adaptation and optimization of the existing geometry is necessary. The
important components of the planned design are shown in \autoref{fig:design}.

A key limitation of the design optimization process is the manual input and
adaption of shapes based on simulations that must be repeated accordingly. An
automation of these steps is desired in order to accelerate and simplify the
design process. This leads to numerical shape optimization. Since the spatial
description of the electric field inside the gun follows (the electrostatic
approximation of) Maxwell's partial differential equations (PDEs), the shape
optimization problem is PDE-constrained \cite{Hinze_2008aa}. Furthermore, there
commonly is no closed-form solution available for complex geometries, so the
PDE is solved numerically, for example, by finite elements \cite{Monk_2003aa}.
PDE-constrained optimization is well known in the computational electromagnetics
community, see the textbook \cite{Di-Barba_2010aa} and references therein.
Particularly in the context of electron guns, several design workflows to
optimize their geometry have been proposed in the last decades
\cite{Lewis_2004ac, Jiang_2015aa, Ribton_2015aa, Stancari_2017aa, Tulu_2018aa}.
However, all of them belong to the class of parameter-based optimization, i.e.,
the designer has to create a template which contains the design variables
describing the geometry, e.g., width, height, and radius. This restricts the
design space and is an inconvenient manual effort. On the other hand, computer
aided design (CAD) tools allow freeform shapes in terms of splines and
non-uniform rational basis splines (NURBS) \cite{deboor1972, Piegl_1997aa}.

\begin{figure}[b]
    \begin{center}
    \includegraphics[width=0.45\textwidth]{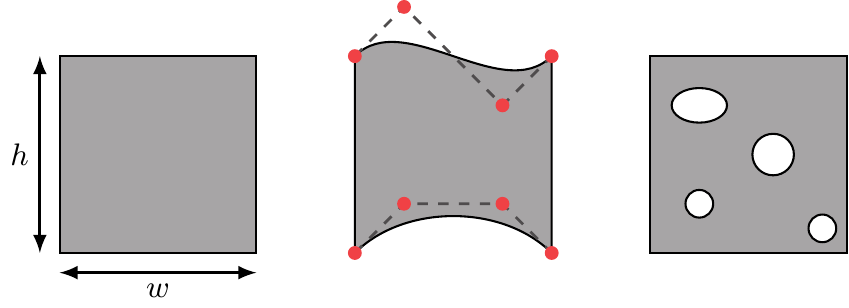}
    \caption{Different types of design optimization. From left to right:
             parameter, shape, and topology optimization.}
    \label{fig:optimization_types}
    \end{center}
\end{figure}

Numerical shape optimization uses the parameters of these NURBS as the degrees
of freedom (DOFs) and thus allows for an improved balance between design
freedom and ease of implementation. Both parameter and shape optimization may
also be used to describe the shape and position of holes in the geometry,
however neither is able to introduce new ones. This requires a further
generalization and leads to topology optimization; this however is not of
interest for our application. An illustration of the different types of
geometric design optimization is given in \autoref{fig:optimization_types}.

There are additional important differences between our method and previous
workflows. In the approaches cited above, each geometry realization is
discretized separately, which requires rather fine spatial resolutions to avoid
numerical errors due to remeshing (`mesh noise') and may again require
additional manual intervention. According to Sandia Labs about
\SI{75}{\percent} of the simulation time in research laboratories is spent on
modeling, parameterization, mesh generation, as well as pre- and post-processing
\cite{Boggs_2005aa}. Yet another distinction should be made regarding the
quality of the resulting field solutions. The current state of the art in the
accelerator community are low order finite element codes, e.g.,
\textsc{POISSON} \cite{halbach1976}. However, even higher-order classical
finite element codes, such as CST \cite{CST_2019aa} yield noisy fields due to
the lack of global regularity, see \cite[Figure 4]{Gjonaj_2009aa}. This is
cumbersome for particle tracking and either needs smoothing or dedicated
(symmetry preserving, mixed element) meshing. To avoid these problems, this
paper proposes a spline-based shape optimization workflow using isogeometric
analysis (IGA) \cite{Cottrell_2009aa}, which integrates finite element analysis
into the conventional NURBS-based CAD design workflow and allows for integrated
particle tracking. IGA-based optimization is well established in many
communities, but less explored in electromagnetism. However,
\cite{Nguyen_2014ab, Pels_2015aa} applied IGA-based optimization to accelerator
magnets (without tracking) and more recently \cite{Merkel_2019aa} suggested a
freeform optimization workflow based on shape calculus for rotating electric
machines; more references can be found in the survey article
\cite{Bontinck_2017ag}.

The paper is structured as follows: after this introduction,
\autoref{sec:splines} gives a short summary on CAD geometry handling and
introduces splines. The following \autoref{sec:field} introduces the electric
field problem, its weak formulation and discretization. Then
\autoref{sec:shape} formulates the optimization problem and introduces
numerical methods for its solution, and \autoref{sec:numerical} discusses the
results for the particular gun in the context of \mbox{Photo-CATCH}. Finally,
the paper closes with conclusions and an outlook.

\section{Splines and geometry} \label{sec:splines}
CAD models are essentially represented by B-splines \cite{deboor1972} and NURBS
\cite{Piegl_1987aa}, since they can exactly describe circular objects, allow
local smoothness control, and give an intuitive definition of freeform curves
and surfaces by so-called control points \cite{Cohen_2001aa}.

\subsection{B-splines}
A basis $\{ B_{i,p} \}_{i=1}^{N_1}$ of a one-dimensional B-spline space
$\mathbb{S}^{p}_{\alpha}$ of degree $p$ and regularity $\alpha$ may be
constructed from a knot vector $\boldsymbol{\Xi} = (\xi_1, \xi_2, \dots,
\xi_{n}) \in [0,1]^n$, $\xi_1 \leq \xi_2 \leq \dots \leq \xi_n$ using the
Cox-de Boor algorithm \cite{de-Boor_2001aa}
\begin{align*}
    B_{i,0}(\xi) &= \begin{cases}
        1 \quad \mathrm{if} \quad \xi_i \leq \xi < \xi_{i+1}\\
        0 \quad \mathrm{otherwise}
    \end{cases}\\
    B_{i,p}(\xi) &= \frac{\xi - \xi_i}{\xi_{i+p} - \xi_i} B_{i,p-1}(\xi) +
    \frac{\xi_{i+p+1} - \xi}{\xi_{i+p+1} - \xi_{i+1}} B_{i+1,p-1}(\xi).
\end{align*}
The knot vector uniquely determines the basis and its properties, including
smoothness and the like. The knots need not be unique and the multiplicity
$m_j$ of a knot value $\xi_j$ determines the continuity of the basis in that
knot to be $C^{p-m_j}$. Furthermore, a knot vector is said to be open if its
first and last knot each have multiplicity $p+1$. For geometry modeling this
usually is the case, since it makes the curve interpolatory in these knots. It
also leads to a distinction between the first and last, and the internal knots.
The latter influence the shape of the basis splines, as they represent the
interfaces between each of the polynomial pieces (or elements) that make up the
splines.

\begin{figure*}[t]
    \begin{center}
    \begin{subfigure}{0.32\textwidth}
        \begin{center}
        \includegraphics[width=0.9\textwidth]{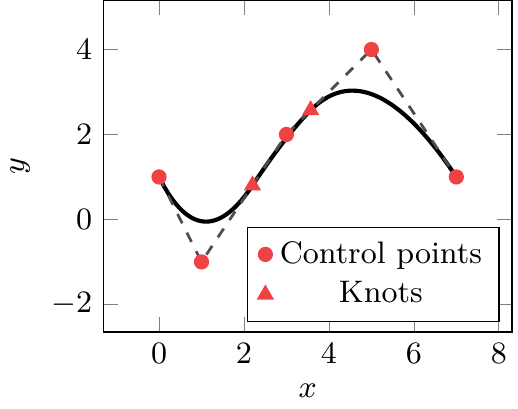}
        \end{center}
    \end{subfigure}
    \hfill
    \begin{subfigure}{0.32\textwidth}
        \begin{center}
        \includegraphics[width=0.9\textwidth]{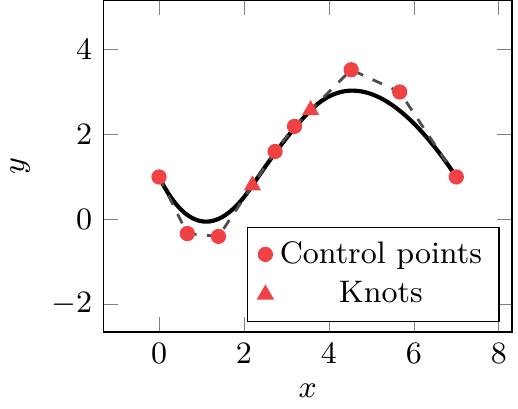}
        \end{center}
    \end{subfigure}
    \hfill
    \begin{subfigure}{0.32\textwidth}
        \begin{center}
        \includegraphics[width=0.9\textwidth]{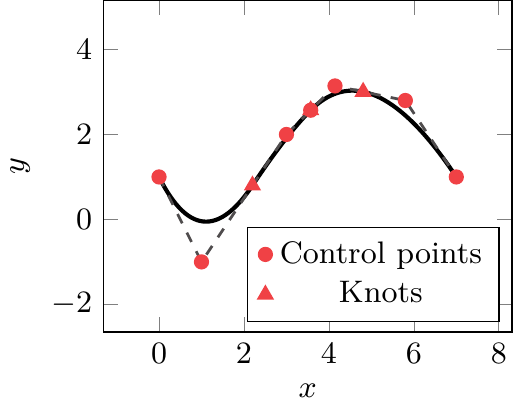}
        \end{center}
    \end{subfigure}

    \begin{subfigure}{0.32\textwidth}
        \begin{center}
        \includegraphics[width=0.9\textwidth]{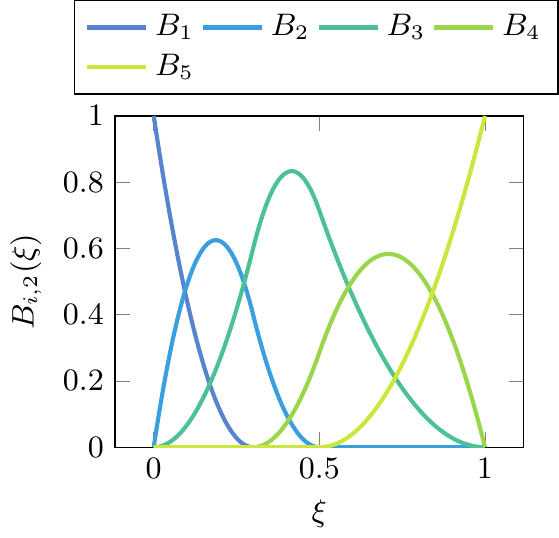}
        \caption{Original curve and basis functions. \hfill \phantom{break
                 without newline}}
        \label{fig:bspline_basis}
        \end{center}
    \end{subfigure}
    \hfill
    \begin{subfigure}{0.32\textwidth}
        \begin{center}
        \includegraphics[width=0.9\textwidth]{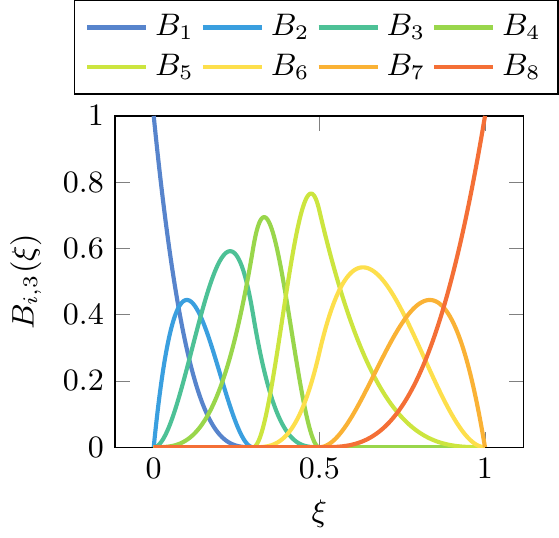}
        \caption{Curve and basis functions after elevating the degree by $1$.}
        \label{fig:bspline_basis_degelev}
        \end{center}
    \end{subfigure}
    \hfill
    \begin{subfigure}{0.32\textwidth}
        \begin{center}
        \includegraphics[width=0.9\textwidth]{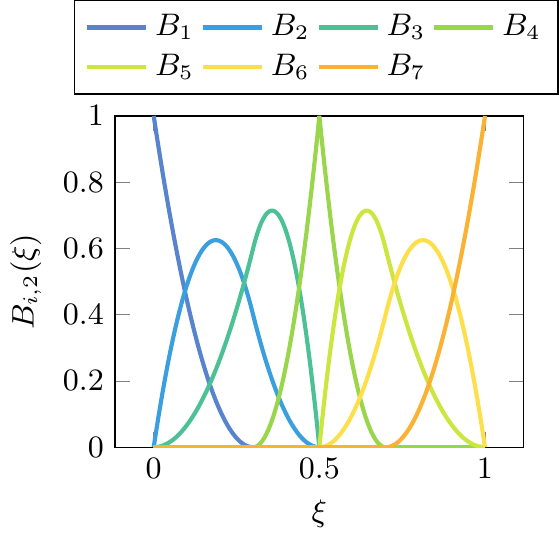}
        \caption{Curve and basis functions after inserting knots at $0.5$ and
                 $0.7$.}
        \label{fig:bspline_basis_kntins}
        \end{center}
    \end{subfigure}
    \caption{Exemplary B-spline curves and their basis functions. The original
             knot vector is $\boldsymbol{\Xi} = (\mathbf{0}_{1 \times 3}, 0.3,
             0.5, \mathbf{1}_{1 \times 3})$ and the control points are
             $\mathbf{P}_1 = (0, 1)$, $\mathbf{P}_2 = (1, -1)$, $\mathbf{P}_3 =
             (3, 2)$, $\mathbf{P}_4 = (5, 4)$ and $\mathbf{P}_5 = (7, 1)$.}
    \label{fig:bsplines}
    \end{center}
\end{figure*}

\subsection{Geometry description}
Given a set of control points $\{ \mathbf{P}_i \}_{i=1}^{N_1} \subset
\mathbb{R}^3$, a three-dimensional B-spline curve is described by a linear
combination of the basis functions
\begin{align}
    C_\mathbf{P}(\xi) = \sum_{i=1}^{N_1} \mathbf{P}_i B_{i,p}(\xi).
    \label{eq:bspline_curve}
\end{align}
This representation is convenient for shape optimization for multiple reasons.
For one, the uniqueness of the basis for a given knot vector leads to an
interpretation of the control points as giving the curve its shape. As a
consequence, changes in the coordinates of the control points directly
translate to changes in the shape of the curve and most importantly they do so
smoothly. An exemplary curve along with the corresponding basis is shown in
\autoref{fig:bspline_basis}. The extension of \eqref{eq:bspline_curve} to the
bivariate case follows from choosing bases $\{ B_{i,p_1} \}_{i=1}^{N_1}$,
$\{ B_{j,p_2} \}_{j=1}^{N_2}$ of $\mathbb{S}^{p_1, p_2}_{\alpha_1, \alpha_2}$
and a control mesh, given by an ordered set of $N_1 \times N_2$ control points
$\mathbf{P}_{i,j}$. A B-spline surface $S_\mathbf{P}$ is then defined via
\begin{align*}
    S_\mathbf{P} = \sum_{i=1}^{N_1} \sum_{j=1}^{N_2} \mathbf{P}_{i,j} B_{i,p_1}
    B_{j,p_2},
\end{align*}
and also volumetric (trivariate) mappings $V_\mathbf{P}$ can be defined
analogously. For the construction and handling of the bi- and trivariate
geometry descriptions we make use of the free \textsc{NURBS} package
\cite{nurbs}.

\subsection{Refinement}
There are several approaches to refine an existing B-spline basis $\{ B_{i,p}
\}_{i=1}^{N_1}$. One is degree elevation, whereby the polynomial degree $p$ of
the basis functions is increased. To preserve the continuity of the original
curve, the multiplicity of each knot ends up being increased alongside the
degree. Furthermore each element, i.e., each polynomial piece, gains new
control points equal to the increase in degree and the positions of all control
points are recomputed such that the shape and parameterization of the curve are
maintained. \autoref{fig:bspline_basis_degelev} shows an example of the process.

A second refinement strategy is given by knot insertion. Here, an arbitrary
internal knot is added to the knot vector. This does not impact the degree of
the basis, however the total number of basis functions is still increased and
the continuity of the basis is reduced in the new knot. For each inserted knot
a new control point is added as well and again the positions of all control
points are determined in a way to keep the shape of the curve intact. An
illustration of the process is given in \autoref{fig:bspline_basis_kntins}.
Note that after inserting a knot at $0.5$ the basis becomes $C^0$ continuous in
that point, as expected. For a more comprehensive treatment of the geometry
descriptions and refinement strategies we refer to \cite{Cottrell_2009aa}.

\section{Field formulation and discretization} \label{sec:field}
\begin{figure}[t]
    \begin{center}
    \includegraphics[width=0.45\textwidth]{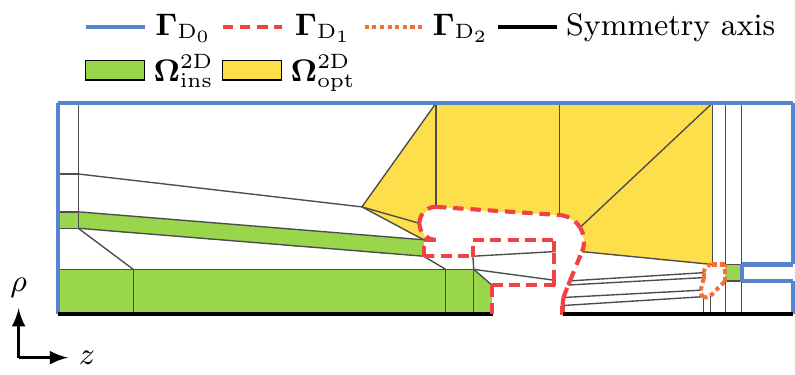}
    \caption{Original geometry and boundary conditions of the domain
             $\boldsymbol{\Omega}^\mathrm{2D}$. Grey lines indicate patch
             boundaries.}
    \label{fig:geometry_orig}
    \end{center}
\end{figure}

Let $\boldsymbol{\Omega}$ be the computational domain of the electron gun with
boundary $\partial \boldsymbol{\Omega}$. In the absence of space charges, the
electric field strength $\mathbf{E} = \mathbf{E}(\mathbf{x}), \mathbf{x} \in
\boldsymbol{\Omega}$ within the gun is described by the electrostatic subset of
Maxwell's equations~\cite{Jackson_1998aa}
\begin{align*}
    \nabla \times \mathbf{E} = 0 \quad \text{and} \quad \nabla \cdot
    (\varepsilon \mathbf{E}) = 0
\end{align*}
in $\boldsymbol{\Omega}$, where the permittivity is given by
\begin{align*}
    \varepsilon = \begin{cases} \begin{alignedat}{2} &\varepsilon_\mathrm{ins}
    &\quad &\mathrm{in}\ \boldsymbol{\Omega}_\mathrm{ins}\\
    & \varepsilon_0 &\quad &\mathrm{otherwise}, \end{alignedat} \end{cases}
\end{align*}
compare \autoref{fig:geometry_orig}. Here $\varepsilon_0$ and
$\varepsilon_\mathrm{ins}$ are the permittivities of empty space and the
insulator respectively. We assume that the domain is given by a multipatch
spline mapping from the reference domain $\boldsymbol{\hat{\Omega}} = (0, 1)^3$
to the physical domain, that is, $\boldsymbol{\Omega} =
\boldsymbol{\Omega}(\mathbf{P})$ in terms of control points $\mathbf{P}$, see
\cite{Buffa_2015aa}. Introducing the electric scalar potential $\phi$ by
$\mathbf{E} = - \nabla \phi$ yields the boundary value problem
\cite[Section 1.7]{Jackson_1998aa}
\begin{align}
    \nabla \cdot (\varepsilon \nabla \phi) &= 0 \quad \mathrm{in}\
    \boldsymbol{\Omega}, \label{eq:estatics}
\end{align}
along with the Dirichlet boundary conditions $\phi = \phi_\mathrm{D_i}$ on
$\boldsymbol{\Gamma}_\mathrm{D_i}$ ($i=0, 1, 2$), see
\autoref{fig:geometry_orig}.

\subsection{Weak formulation}
Exploiting the axisymmetry of the configuration we may restrict our analysis to
$\boldsymbol{\Omega}^\mathrm{2D}$, i.e., the $\rho$-$z$-plane. Let $V =
H^1(\boldsymbol{\Omega}^\mathrm{2D})$ denote the space of square-integrable
functions with square-integrable gradients \cite{Monk_2003aa}. Following the
Ritz-Galerkin approach, we deduce the weak form of \eqref{eq:estatics} as:
find ${\phi} \in V_D$ such that
\begin{align}
    \int_{\boldsymbol{\Omega}^\mathrm{2D}} \varepsilon \nabla \phi \cdot \nabla
    {\phi}'\ \rho\: \mathrm{d}\rho\: \mathrm{dz} = 0 \label{eq:variational}
\end{align}
for every $\phi' \in V_0$, where $V_D$ indicates the space of functions in $V$
satisfying the Dirichlet boundary conditions, while $V_0$ indicates the $V$
subspace of functions vanishing on $\boldsymbol{\Gamma}_\mathrm{D}$. A
finite-element-like discretization of \eqref{eq:variational} is obtained by
restricting to a finite-dimensional subspace $V_h \subset V$. Using basis
functions $\{ v_i \}_{i=1}^{N}$ of $V_h$ we express the potential as
\begin{align}
    \phi_h &= \sum_{i=1}^{N} \varphi_i v_i, \quad \varphi_i \in \mathbb{R}
    \label{eq:phi_h}
\end{align}
and the approximated electric field strength follows from $\mathbf{E}_h = -
\nabla\phi_h$. The linear system of equations reads
\begin{align}
    \mathbf{K}_\varepsilon \boldsymbol{{\varphi}} &= -\boldsymbol{\varrho},
    \label{eq:discrete}
\end{align}
where we only consider the $N_\mathrm{dof}$ unconstrained coefficients as
degrees of freedom, i.e.,
\begin{align}
    (\mathbf{K}_\varepsilon)_{ij} &= \int_{\boldsymbol{\Omega}^\mathrm{2D}}
    \varepsilon \nabla v_j \cdot \nabla v_i\ \rho\: \mathrm{d}\rho\:
    \mathrm{dz} \label{eq:integral_K}
\end{align}
for $1 \leq i,j \leq N_\mathrm{dof}$ and include the coefficients known due to
boundary conditions in the right-hand side $\boldsymbol{\varrho}$.

Please note, until now the basis functions have not yet been specified. The
next section will propose to use B-splines instead of the more common
finite-element-type hat functions \cite{Cottrell_2009aa}.

\subsection{Isogeometric analysis} \label{subsec:isogeometric}
The main idea of IGA is to use B-splines or NURBS not only for the geometry
description but also to represent the solution. This enables the solution of
numerical problems, as the one defined above, on computational domains without
introducing a geometric modeling error. Moreover, the thereby obtained geometry
parameterization lends itself very nicely towards shape optimization, since it
offers an intuitive set of degrees of freedom which immediately deform the
underlying mesh. Finally, the use of high order B-spline basis functions in
\eqref{eq:discrete} guarantees rapid convergence and a high continuity of the
solution~\cite{Cottrell_2009aa}.

Let $S_\mathbf{P}: \boldsymbol{\hat{\Omega}}^\mathrm{2D} \!\to
\boldsymbol{\Omega}^\mathrm{2D}$ denote the bivariate geometry mapping from the
reference domain $\boldsymbol{\hat{\Omega}}^\mathrm{2D} \!:= (0, 1)^2$ to the
physical domain, from which eventually a 3D description is obtained by
revolution \cite{Simona_2020aa}. Assuming that $S_\mathbf{P}$ is piecewise
smoothly invertible, we may define the approximation space $V_h$ using a
gradient preserving transformation
\begin{align*}
    V_h = \{v: v = \hat{v} \circ S_\mathbf{P}^{-1},\ \hat{v} \in \hat{V}_h \}.
\end{align*}
Here $\hat{V}_h$ is a discrete space on the parametric domain, for which we
elect to use the space of B-splines with degree $p_i$ and continuity $\alpha_i$
along dimension $i$, denoted by $\mathbb{S}^{p_1, p_2}_{\alpha_1, \alpha_2}$ in
the presented two-dimensional case. We have implemented the weak form and the
discretization by IGA within the free software \textsc{GeoPDEs} \cite{geopdes}.

\section{Shape optimization} \label{sec:shape}
The overall aim is to optimize the geometry of the electron gun to achieve two,
possibly competing, goals. Firstly, we want to minimize the maximum electric
field strength on the electrode surface, and secondly, we want to ensure a
proper beam. Since the regions of the geometry that are relevant for field
emission are far away from the cathode and the beam axis, we investigate the
two problems separately. We optimize the shape first and then perform particle
tracking to determine the beam properties of the optimized geometry.

\begin{figure*}[t]
    \begin{center}
        \begin{subfigure}{0.49\textwidth}
            \begin{center}
            \includegraphics[width=0.95\textwidth]{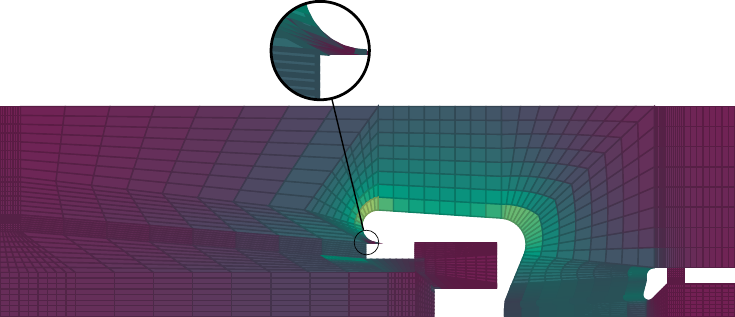}
            \caption{}
            \label{fig:E_orig}
            \end{center}
        \end{subfigure}
        \begin{subfigure}{0.49\textwidth}
            \begin{center}
            \includegraphics[width=0.95\textwidth]{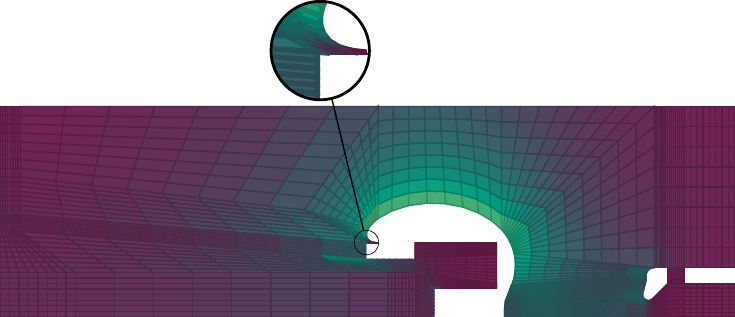}
            \caption{}
            \label{fig:E_cobyla}
            \end{center}
        \end{subfigure}

        \begin{subfigure}{0.49\textwidth}
            \begin{center}
            \includegraphics[width=0.95\textwidth]{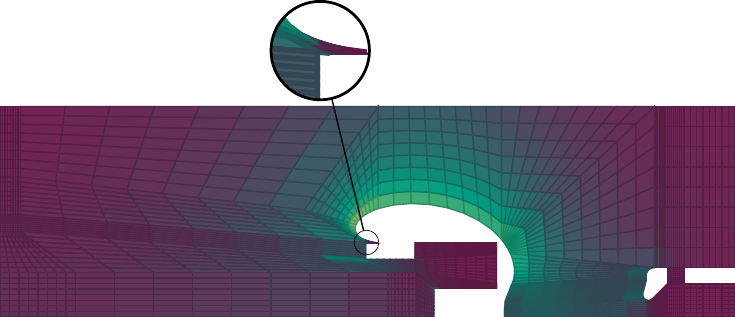}
            \caption{}
            \label{fig:E_tp}
            \end{center}
        \end{subfigure}
        \begin{subfigure}{0.49\textwidth}
            \begin{center}
            \vspace*{.3cm} \hspace*{.75cm}
            \includegraphics[width=1.9\textwidth]{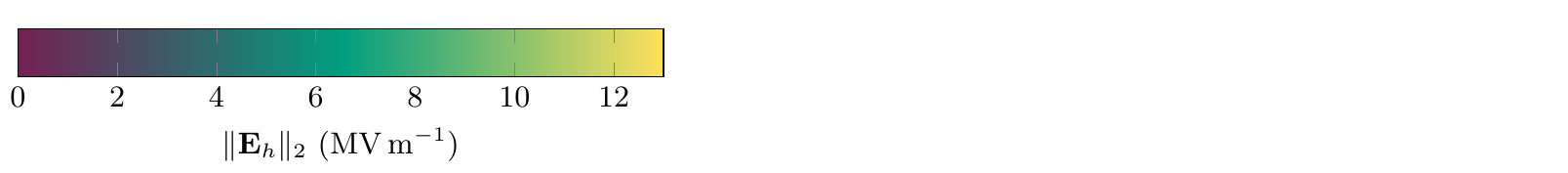}
            \end{center}
        \end{subfigure}
        \caption{Electric field magnitude for the original
                 (\autoref{fig:E_orig}) and optimized (\autoref{fig:E_cobyla}
                 according to \eqref{eq:optimization} and \autoref{fig:E_tp}
                 according to \eqref{eq:tp}) geometries. The plot
                 representation uniformly divides each patch into 8 elements per
                 coordinate direction ($n_\mathrm{sub}=8$) and the pop-outs
                 zoom in on the triple point. Computed using \textsc{GeoPDEs}.}
        \label{fig:E_iga}
    \end{center}
\end{figure*}

Only the shape of the electrode is relevant for the geometry, i.e., the
boundary $\boldsymbol{\Gamma}_{\mathrm{D}_1}$ in \autoref{fig:geometry_orig}.
Furthermore, as can be seen in \autoref{fig:E_orig}, it makes sense to restrict
our attention to the domain $\boldsymbol{\Omega}_\mathrm{opt}^\mathrm{2D}$, as
indicated in \autoref{fig:geometry_orig}. The degrees of freedom for the
optimization are given by the positions of the control points $\mathbf{P}$ of
the curve $C_\mathbf{P}(\rho, z)$ describing that part of
$\boldsymbol{\Gamma}_{\mathrm{D}_1}$, which intersects with
$\boldsymbol{\Omega}_\mathrm{opt}^\mathrm{2D}$. On a further note, the volume
of the electrode may not exceed some fixed value $V_\mathrm{c}$ due to space
and weight considerations. For this, let $V_\mathrm{el}(\mathbf{P})$ denote the
volume of the electrode in dependence on
$\boldsymbol{\Gamma}_{\mathrm{D}_1}(\mathbf{P})$, as characterized in
\autoref{fig:design}.

We allow geometries from an admissible set
\begin{align*}
    \mathcal{A} = \left\{
    (\mathbf{P}_1, \dotsc, \mathbf{P}_{N_\mathrm{opt}}): \underline{\mathbf{P}
    }_i \leq \mathbf{P}_i \leq \overline{\mathbf{P}}_i,\ i=1, \dotsc,
    N_\mathrm{opt} \right\},
\end{align*}
where $\leq$ is to be read componentwise. $\mathcal{A}$ accounts for
constraints on the coordinates of the control points in terms of upper
$\overline{\mathbf{P}}_i$, and lower bounds $\underline{\mathbf{P}}_i$, for
example, to avoid intersections. The optimization problem is finally obtained as
\begin{align}
    \min_{\mathbf{P} \in \mathcal{A}} \max_{\mathbf{x} \in
    \boldsymbol{\Omega}_\mathrm{opt}^\mathrm{2D}(\mathbf{P})} \|
    \mathbf{E}_h(\mathbf{x}; \mathbf{P}) \|_2 \label{eq:optimization}
\end{align}
subject to
\begin{align}
    \mathbf{E}_h(\mathbf{x}; \mathbf{P}) &= - \nabla
    \phi_h(\mathbf{x}; \mathbf{P}) \quad \mathrm{via\ (\ref{eq:phi_h}
    -\ref{eq:discrete})} \nonumber\\
    V_\mathrm{el}(\mathbf{P}) &\leq V_\mathrm{c} \nonumber\\
    \mathbf{f}_\mathrm{track}(\mathbf{E}_h(\mathbf{x}; \mathbf{P})) &<
    \textbf{tol}, \label{eq:tracking_constraint}
\end{align}
where $\mathbf{x} = (\rho, z)^\top$ is a position in the $\rho$-$z$-plane and
the inner optimization, $\max\| \mathbf{E}_h(\mathbf{x}; \mathbf{P}) \|_2$, is
approximated by a discrete maximum over a set of sample points which are used
for the numerical quadrature of \eqref{eq:integral_K}. The function
$\mathbf{f}_\mathrm{track}$ denotes quantities of interest from the particle
tracking, as defined in \autoref{sec:results}, and $\textbf{tol}$ describes
associated bounds that ensure functionality.

The above formulation of the optimization problem neglects two aspects which
may also be critical for the electrostatic design of electron guns: the high-
voltage cable conductor and the electric field magnitude at the triple point,
where electrode, insulator and vacuum meet. The position of the triple point is
highlighted in \autoref{fig:E_iga} by the pop-outs. Including the high-voltage
cable will have little effect on the maximum field on the electrode surface,
but it may significantly influence the field at the triple point. To showcase
the flexibility and effectiveness of our approach, we also add a term to
\eqref{eq:optimization} that aims to minimize the field at the triple point
$\mathbf{x}_\mathrm{tp}$ leading to the objective function
\begin{align}
    \min_{\mathbf{P} \in \mathcal{A}} \bigg( \max_{\mathbf{x} \in
    \boldsymbol{\Omega}_\mathrm{opt}^\mathrm{2D}(\mathbf{P})} \|
    \mathbf{E}_h(\mathbf{x}; \mathbf{P}) \|_2 + w \sum_{i=1}^{N_\mathbf{U}} \|
    \mathbf{E}_h(\mathbf{x}_i; \mathbf{P}) \|_2 \bigg), \label{eq:tp}
\end{align}
where $w$ is a weighting factor to balance the two terms and $\mathbf{U} =
\mathbf{U}(\mathbf{x}_\mathrm{tp})$ is a neighborhood of
$\mathbf{x}_\mathrm{tp}$, from which $N_\mathbf{U}$ sample points are taken to
approximate the value at the triple point. It should be noted that the field
becomes infinite at the triple point due to the sharp corner of the geometry,
see \autoref{fig:geometry_orig} and \cite[Section 2.11]{Jackson_1998aa}.
Therefore, we only evaluate the discrete representation of the field in the
sample points, such that this issue can be mitigated. The results in
\autoref{sec:numerical} indicate that this improves the design, but the results
must still be treated with care.

The question of which optimization algorithm to employ for solving the given
problem is determined by the lack of smoothness of the $\min \max$ problem, the
unavailability of derivatives, and the nature of the constraints. In this work,
we use a two step process consisting of the successive application of a global,
followed by a local, optimization algorithm. The global algorithm
(\textsc{ISRES}) is an evolution strategy based on a stochastic ranking to
balance the objective function with a constraint based penalty function
\cite{runarsson2005}. It does not need to compute or estimate derivatives of
the objective function and at the same time is able to handle arbitrary
nonlinear constraints, thus it meets our requirements. However the associated
computational effort is comparatively high, as is to be expected with global
optimization in general and evolutionary algorithms in particular. The local
algorithm (\textsc{COBYLA}) works by creating linear approximations of both the
objective and constraint functions via interpolating their evaluations at the
vertices of a simplex \cite{powell1994}. This method again meets our criteria
of not needing to compute derivatives of the objective function and being able
to deal with nonlinear constraints. Even more importantly, the computational
cost of this local algorithm is much less compared to that of the global one.
We can therefore select a smaller tolerance on the change of the objective
function over consecutive iterations and still obtain results in a shorter
period of time.

For either algorithm we make use of the freely available implementations from
the \textsc{NLopt} package \cite{nlopt}. Derivatives, alternative formulations,
or approximations of the optimization problem \eqref{eq:optimization} may allow
for more sophisticated algorithms. In particular, one may look for a `smoother'
objective function that avoids the discrete maximum, or aim for convexity of
the optimization problem.

\section{Numerical results} \label{sec:numerical}
\begin{figure}[t]
    \begin{center}
    \includegraphics[width=0.4\textwidth]{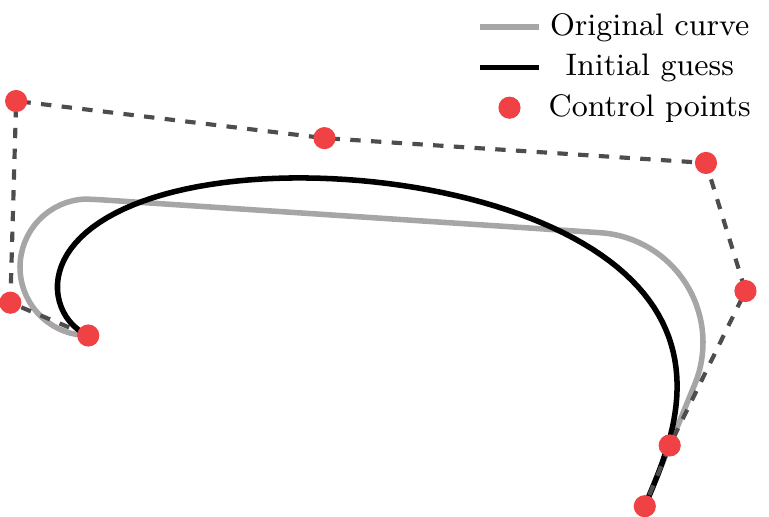}
    \caption{Original curve and the least squares fit serving as the initial
             shape for the optimization.}
    \label{fig:curve_orig}
    \end{center}
\end{figure}

Based on the abstract formulation of the optimization problem given in
\eqref{eq:optimization}, let us discuss the specific choices for the electron
gun shown in \autoref{fig:design}. We begin the optimization procedure with a
B-spline curve of degree $p=7$ without any internal knots. This is equivalent
to a simple polynomial of degree $7$, however both in terms of the presented
isogeometric setting, and also for later refinements, it makes sense to
interpret it in the B-spline context. The curve is a reasonable compromise
between design freedom, simplicity, and the desire to obtain a smooth and
manufacturable solution. The control points of the initial guess are determined
by a least squares fit of the original `flat' design, as shown in
\autoref{fig:curve_orig}, and the exact parameters can be found in
\cite{egunopt}. In order to keep the overall geometry intact, the first and
last control points are both fixed in their original positions over the course
of the optimization.

\subsection{Optimization results}
Two successive optimization cycles are performed, as described in
\autoref{sec:shape}. The first one uses a global optimization algorithm
(\textsc{ISRES}) with a relative tolerance of $10^{-3}$ on the objective
function, and the second utilizes a local algorithm (\textsc{COBYLA}) with a
relative tolerance of $10^{-4}$. The volume constraint is set at $V_\mathrm{c}
= \SI{625}{\centi\meter\cubed}$, based on the assumptions that the insulator
assembly can support a maximum weight of \SI{5}{\kilo\gram} and a stainless
steel (type: 1.4404) electrode is used. The bounds for the admissible set can
be found in \cite{egunopt}.

\begin{figure}[b]
    \begin{center}
    \hspace*{.3cm} \includegraphics[width=0.55\textwidth]{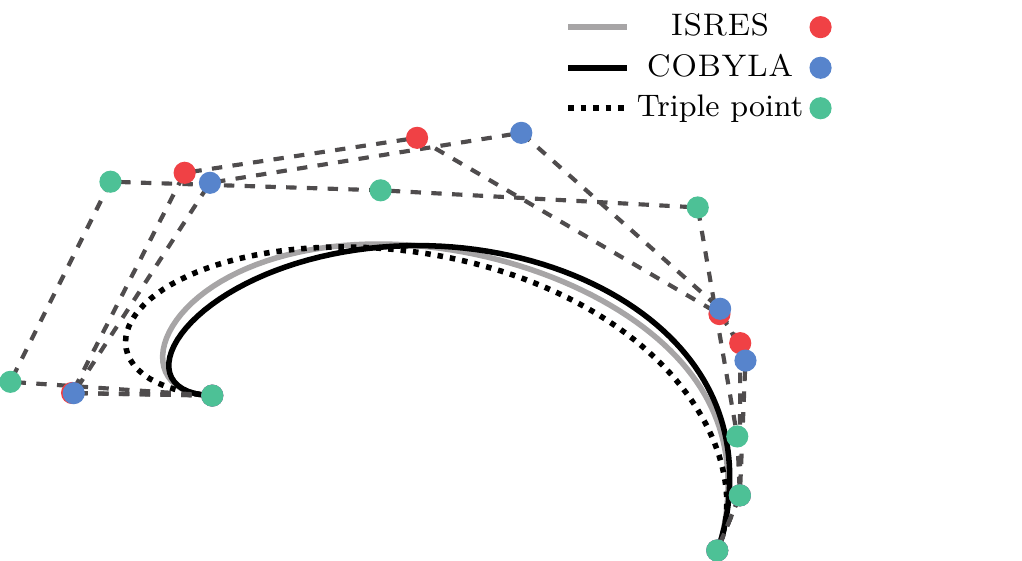}
    \caption{Curves obtained from optimizations employing \textsc{ISRES}, \textsc{COBYLA} and the modified objective function \eqref{eq:tp} respectively.}
    \label{fig:curve_opt}
    \end{center}
\end{figure}

The resulting shapes are shown in
\autoref{fig:curve_opt} and for comparison, we also include the optimized curve
obtained by using the modified objective function \eqref{eq:tp}. As the
high-voltage cable is still missing from the model, we only use the local
algorithm for the second formulation, since the results are most likely not
reliable for the final design. Nonetheless we find that the shapes look similar
and the larger bulge at the back of the electrode for the modified objective is
suitable for shielding the triple point. From this point onward we refer to the
curve obtained via \textsc{COBYLA}, compare \autoref{fig:curve_opt}, as the
optimized shape if not explicitly stated otherwise and it will serve as the
starting point for further analyses. Regarding the computational effort, the
global optimization algorithm took about a week to find a solution satisfying
our strict numerical tolerances, however it is possible to lower this number
significantly by choosing a larger value for the tolerance or electing to only
perform a local optimization. In contrast, the local algorithm only required
computation times of around 7 hours to find a sufficiently accurate solution.

The electric field solutions corresponding to the original and both optimized
curves are depicted in \autoref{fig:E_iga}. We observe a clearly visible
reduction of the maximum field strength, and in addition, the change in the
electric field magnitude along the electrode appears to be smoother for the optimized
geometries. For the solution based on the isogeometric technique described in
\autoref{subsec:isogeometric}, the open source package \textsc{GeoPDEs} is used
\cite{geopdes}. The B-spline space is chosen as $\mathbb{S}^{3,3}_{2,2}$ and
each of the parametric domains, of the patches indicated in
\autoref{fig:geometry_orig}, is divided into $n_\mathrm{sub} = 16$ elements per
coordinate direction using uniform knot insertion. For verification, both the
original and optimized geometries are imported into CST Studio Suite 2019 and
the field problem is solved using their adaptive mesh refinement with a
tolerance of $10^{-4}$, based on a discretization with second order tetrahedral
elements. The Dirichlet boundary conditions, as marked in
\autoref{fig:geometry_orig}, are chosen as $\boldsymbol{\Gamma}_\mathrm{D_0} =
\SI{0}{\volt}$, $\boldsymbol{\Gamma}_\mathrm{D_1}(\mathbf{P}) = \SI{
-300}{\kilo\volt}$, and $\boldsymbol{\Gamma}_\mathrm{D_2} = \SI{1}{\kilo\volt}$.

\begin{figure}[t]
    \begin{center}
    \includegraphics[width=0.45\textwidth]{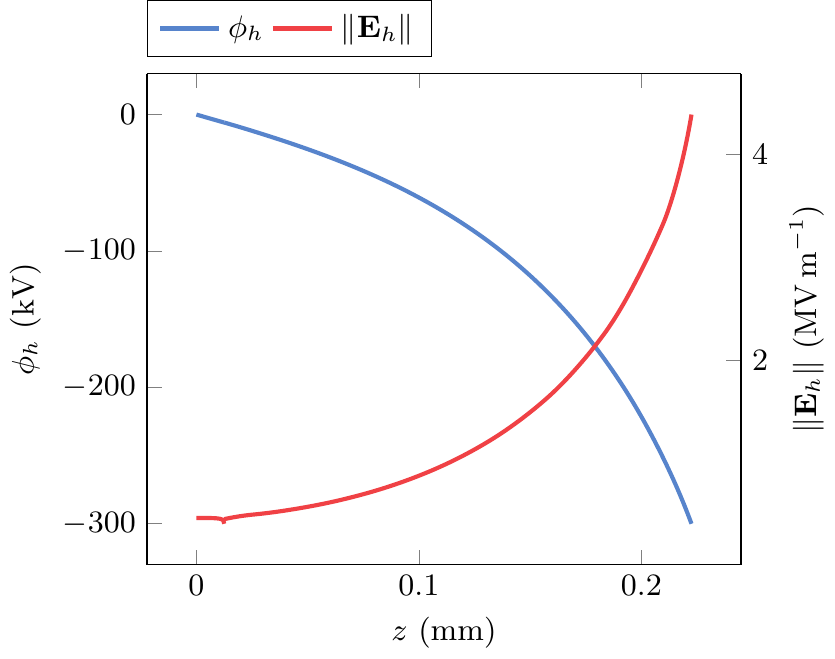}
    \caption{Potential distribution and magnitude of the electric field
             gradient along the outer insulator.}
    \label{fig:insulator_potential}
    \end{center}
\end{figure}

The numerical values of the objective functions and the volume constraint, for
the original and both optimized geometries respectively, are listed in
\autoref{tab:results_opt}, where
\begin{align*}
    E_\mathrm{max}^{\star} &= \max_{\mathbf{x} \in
    \boldsymbol{\Omega}_\mathrm{opt}^\mathrm{2D}(\mathbf{P})} \|
    \mathbf{E}_h(\mathbf{x}; \mathbf{P}) \|_2
\end{align*}
is introduced for brevity; $\star$ refers to the used code. We observe a
significant reduction in the maximum electric field strength, such that it falls
well below the desired \SI{10}{\mega\volt\per\meter} for the optimized
electrode. The volume constraint is also fulfilled at
\SI{618}{\centi\meter\cubed} even though the initial shape had violated the
requirement (\SI{630}{\centi\meter\cubed}). Finally, it can be seen that the
results of our code (`IGA') and CST's EM Studio (`CST') are in good agreement.

\begin{table}[h]
    \begin{ruledtabular}
    \caption{Electric field magnitude at critical points for the original and
             optimized geometries. We use $\cdot_\mathrm{tp}$,
             $\cdot_\mathrm{c}$ and $\cdot_\mathrm{ar}$ to refer to the values
             at the triple point, cathode and anode ring respectively.}
    \begin{tabular}{c|ccccc}
    (\si{\mega\volt\per\meter}) & $E_\mathrm{max}^\mathrm{IGA}$ &
    $E_\mathrm{max}^\mathrm{CST}$ & $E_\mathrm{tp}^\mathrm{IGA}$ &
    $E_\mathrm{c}^\mathrm{IGA}$ & $E_\mathrm{ar}^\mathrm{IGA}$ \\
    \hline
    Original & 13 & 12.93 & 2.55 & 2.31 & 6.5 \\
    Optimized & 9.06 & 9.06 & 3.27 & 2.99 & 5.63 \\
    Triple point & 10.94 & - & 1.9 & 3.22 & 5.3
    \end{tabular}
    \label{tab:results_opt}
    \end{ruledtabular}
\end{table}

Apart from the electrode, the maximum electric field strength on the cathode
surface is also of interest. The numerical solution gives a value of
\SI{2.99}{\mega\volt\per\meter} ($E_\mathrm{c}^\mathrm{IGA}$) for the optimized
geometry, well below the \SI{3.9}{\mega\volt\per\meter} that are documented for
the former Jefferson Lab FEL gun that was routinely operated at
\SI{-320}{\kilo\volt} \cite{siggins2001}. This value is expected to yield a
sufficiently low energy spread \cite{friederich2015}, and preliminary results
are shown in \autoref{sec:results}. A more thorough integration of a
particle tracking software into the shape optimization process will allow
further improvement upon this value.

Another important quantity is the magnitude of the electric field at the
so-called triple point, where electrode, insulator, and vacuum meet. Closeups
of the field surrounding the triple point are shown in \autoref{fig:E_iga}. Our
simulations predict a value of \SI{3.27}{\mega\volt\per\meter}
($E_\mathrm{tp}^\mathrm{IGA}$) for the optimized geometry, a significant
increase compared to the \SI{2.55}{\mega\volt\per\meter} for the original
geometry. The studies in \cite{garcia2016} suggest that one should
aim for field strengths below \SI{1}{\mega\volt\per\meter} at the triple point.
We see that the modified formulation \eqref{eq:tp} significantly reduces the
field strength at this point. However, it may be necessary to further optimize
the design and include additional shielding to minimize the field gradient at
this critical point, as shown in \cite{serrano2018}. Such measures could also
influence the potential distribution and electric field magnitude along the
outer insulator at the back of the electrode. Numerical results for these
quantities can be seen in \autoref{fig:insulator_potential} and show a
nonlinear behavior. Adapting the design to linearize the field strength along
the insulator surface may improve performance and reduce the chance of
electrical breakdown at high voltages \cite{garcia2016, garcia2019}.

\begin{figure}[b]
    \begin{center}
    \includegraphics[width=0.45\textwidth]{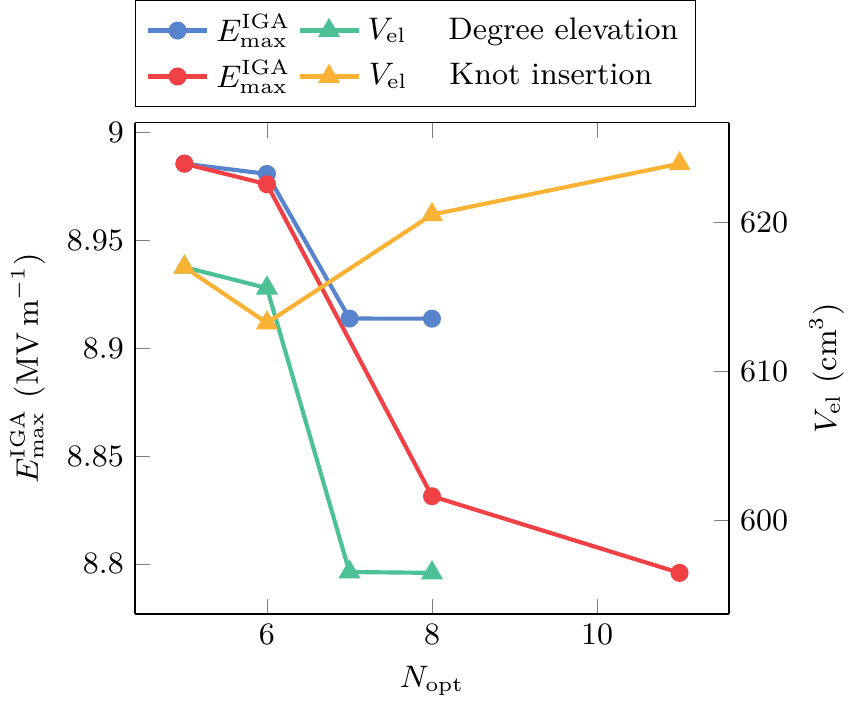}
    \caption{Convergence of maximum electric field magnitude and volume
             constraint with respect to curve degree and number of internal
             knots.}
    \label{fig:cvg_curve}
    \end{center}
\end{figure}

Lastly, the maximum value
of the field gradient on the surface of the anode ring is
\SI{5.63}{\mega\volt\per\meter} ($E_\mathrm{ar}^\mathrm{IGA}$) according to our
computations. It is also possible to further reduce this value, since the shape
of the anode ring was not optimized in this work. The field magnitudes, at all
critical points and for both the original and optimized geometries, are listed
in \autoref{tab:results_opt}. Looking at the results, it can be seen that while
decreasing the field strength at the triple point, the second formulation
\eqref{eq:tp} leads to a higher gradient on the electrode surface.

\begin{figure*}[t]
    \begin{center}
    \begin{subfigure}{0.38\textwidth}
        \begin{center}
        \includegraphics[width=0.9\textwidth]{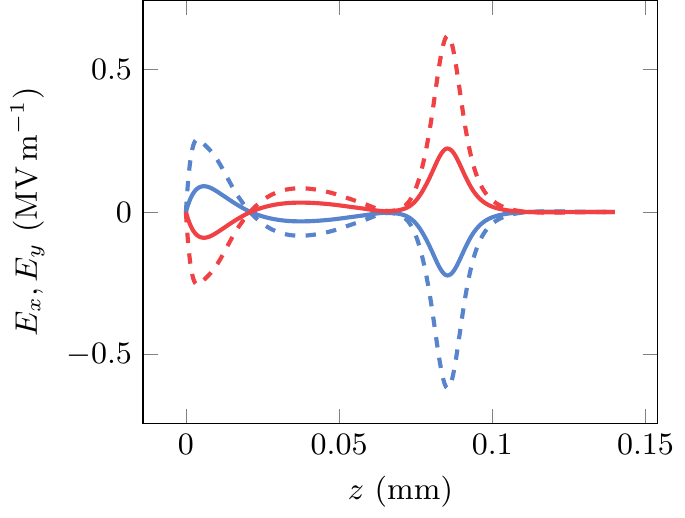}
        \end{center}
    \end{subfigure}
    \hfill
    \begin{subfigure}{0.22\textwidth}
        \begin{center}
        \includegraphics[width=0.9\textwidth]{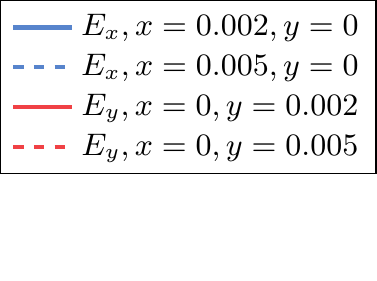}
        \end{center}
    \end{subfigure}
    \hfill
    \begin{subfigure}{0.38\textwidth}
        \begin{center}
        \includegraphics[width=0.9\textwidth]{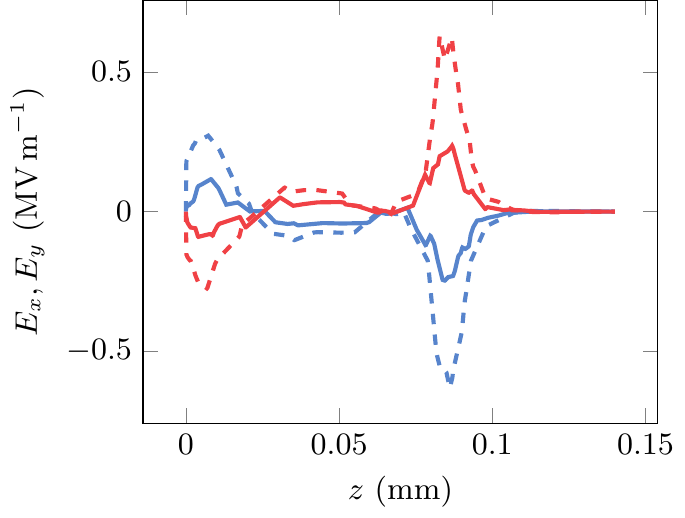}
        \end{center}
    \end{subfigure}
    \end{center}
    \caption{Comparison of electric field solutions computed using IGA (left)
             and linear tetrahedral finite elements (right). The results
             clearly showcase the advantages of IGA for obtaining smooth fields
             for tracking purposes.}
    \label{fig:loworder}
\end{figure*}

We conclude our study of the geometry by looking at the convergence of the
optimized parameters with respect to degree elevation and knot insertion. As
discussed in \autoref{sec:splines}, both refinement types add control points to
an existing curve which increases the number of degrees of freedom. In the case
of knot insertion the solution space is expanded even further, since the
continuity of the basis in the new knot values is reduced, thus allowing a
reduced continuity of the curve. The results in terms of the maximum electric
field strength and the volume of the electrode are shown in
\autoref{fig:cvg_curve}. In the case of degree elevation, the degree of the
curve is continually increased by 1, i.e., $p \in \{ 7, 8, 9, 10 \}$. For knot
insertion, the intervals of the underlying knot vector are repeatedly halved by
inserting additional knots, i.e., $\boldsymbol{\Xi}_0 =
(\mathbf{0}_{1 \times 7}, \mathbf{1}_{1 \times 7})$, $\boldsymbol{\Xi}_1 =
(\mathbf{0}_{1 \times 7}, \frac{1}{2}, \mathbf{1}_{1 \times 7})$,
$\boldsymbol{\Xi}_2 = (\mathbf{0}_{1 \times 7}, \frac{1}{4}, \frac{1}{2},
\frac{3}{4}, \mathbf{1}_{1 \times 7})$, $\boldsymbol{\Xi}_3 =
(\mathbf{0}_{1 \times 7}, \frac{1}{8}, \frac{1}{4}, \frac{3}{8}, \frac{1}{2},
\frac{5}{8}, \frac{3}{4}, \frac{7}{8}, \mathbf{1}_{1 \times 7})$, while the
degree is kept constant at $p=7$. The corresponding optimization cycles are
carried out with \textsc{COBYLA}. One can clearly observe a correlation between
the number of control points $N_\mathrm{opt}$ and the quality of the solution.
For this example, the solutions based on knot insertion seem to make better
use of the available volume when compared to the ones from degree elevation,
however this may simply be due to a local optimum.

\subsection{Smoothness of IGA solutions}
In \autoref{sec:introduction} and \autoref{subsec:isogeometric} we mentioned
the higher global smoothness of the discrete fields when using IGA instead of
classical finite elements. This is especially relevant for tracking
applications, as the quality of the particle trajectories directly depends on
the quality and properties of the electric field. Even if the tracking tool
only supports pointwise data import; this is, for example, the case for
\textsc{ASTRA} \cite{astra}, regularity can be reconstructed if high order
interpolation is used. \textsc{ASTRA} can make use of higher order polynomial
interpolation internally, such that fields and first derivatives with respect
to the space coordinates are continuous functions, e.g., when computing space
charge effects \cite[Sections 4.3, 4.4 and 6.9]{astra}. To illustrate our point,
\autoref{fig:loworder} shows a comparison between an IGA based solution using
\textsc{GeoPDEs} and a linear finite element solution from CST. We use $E_x,
E_y$ to denote the $x$- or $y$-components of the electric field strength
respectively and look at the field close to the beam axis. Again we employ the
adaptive refinement tool of CST with a tolerance of $10^{-4}$ to obtain a
`realistic' comparison. The results show a distinct difference in favor of IGA,
since the unstructured mesh of the FE method, combined with the low order basis
functions, leads to noisy fields.

As the shape optimization is only carried out in
$\boldsymbol{\Omega}_\mathrm{opt}^\mathrm{2D}$, one may assume that the
tracking results are not severely affected by it, and therefore we did not
consider the last constraint \eqref{eq:tracking_constraint} within the
optimization. Nonetheless, the following section shows a preliminary
investigation of the emission process and electron acceleration to ensure a
solid gun performance with the optimized geometry.

\subsection{Particle tracking} \label{sec:results}
Aside from fulfilling the previously discussed optimization criteria, the
electric field should also be suitable for the initial acceleration of
electrons emitted from the photocathode. In order to verify this, we perform
simulations using the well-established particle tracking software
\textsc{ASTRA} \cite{astra}. Once the particle trajectories are computed, it is
possible to evaluate statistical quantities that give insight into the gun
performance. Two of these quantities of interest are the root mean square
(RMS) beam width $\mathbf{x}_\mathrm{rms} \in \mathbb{R}^{N_z}$ and the related
normalized transverse RMS emittance $\boldsymbol{\epsilon}_x \in
\mathbb{R}^{N_z}$, both in the $x$- and $y$- direction, where $N_z$ is the
number of discrete points along the $z$-axis where the trajectories are known.

\begin{figure}[t]
    \begin{center}
    \includegraphics[width=0.4\textwidth]{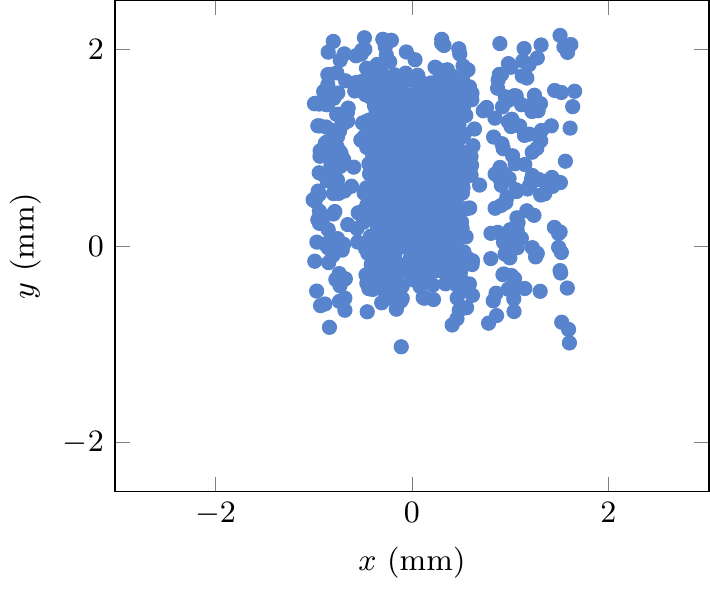}
    \caption{Initial spatial distribution of the $N_\mathrm{p} = 2^{11}$
             macroparticles to be emitted from the cathode. The positions are
             sampled from a measurement of the diode laser in use at
             \mbox{Photo-CATCH}.}
    \label{fig:astra_init_space}
    \end{center}
\end{figure}

In this context, \autoref{fig:astra_init_space} shows the initial macroparticle
distribution in space for $N_\mathrm{p} = 2^{11}$ particles, which was
obtained from a measurement using the DataRay BeamMap2 Beam Profiler. The laser
spot has an oval shape with RMS radii of $r_{x,\mathrm{rms}} =
\SI{0.41}{\milli\meter}$ and $r_{y,\mathrm{rms}} = \SI{0.72}{\milli\meter}$.
The corresponding data is also available at \cite{egunopt}. The emission times
of the particles are drawn from a normal distribution with mean \SI{0}{\second}
and standard deviation \SI{5}{\pico\second}. This represents a practical
compromise between accuracy and simplicity, however there exists extensive work
on the details of the emission process and the bunch profile in time
\cite{espig2014}. The thermal emittance, representing the minimal emittance of
a photoemission electron source, depends on material properties and the
illuminating wavelength \cite{bazarov2008}, and therefore on the photocathode
material. In order to conduct a more general simulation, we thus assume the
particles to have no initial momentum. The total bunch charge is estimated at
\SI{100}{\femto\coulomb}, corresponding to the planned operation of the gun at
\mbox{Photo-CATCH}. The gun is expected to produce a continuous waveform beam
with a current of \SI{300}{\micro\ampere} at a repetition rate of
\SI{3}{\giga\hertz}, which is optimized for the operational parameters of the
superconducting electron accelerator \mbox{S-DALINAC} at TU Darmstadt
\cite{richter1996, pietralla2018}.

\begin{table}[h]
    \begin{ruledtabular}
    \caption{Relative errors in the quantities of interest; for the actual
             values see \autoref{tab:results_qoi}.}
    \begin{tabular}{c|cccccc}
    (\si{\percent}) & $\delta_{x_\mathrm{rms}}$ & $\delta_{y_\mathrm{rms}}$ & $\delta_{z_\mathrm{rms}}$ & $\delta_{\epsilon_x}$ & $\delta_{\epsilon_y}$ & $\delta_{\epsilon_z}$\\
    \hline
    Initial & 4.563 & 2.243 & 0.942 & 8.423 & 10.439 & 6.262 \\
    Refined & 1.553 & 0.992 & 0.514 & 4.079 & 4.305 & 3.437 \\
    \end{tabular}
    \label{tab:results_astra}
    \end{ruledtabular}
\end{table}

In addition to the already described parameters we initially choose a total of
$N_\mathrm{p} = 2^{11}$ macroparticles, a time step of about $\Delta \mathrm{t}
= \SI{0.244}{\pico\second}$ for the Runge-Kutta integrator, and we set the grid
for the electric field to be equidistant with $n_x = n_y = 16$ points ($\Delta
x = \Delta y = \SI{0.156}{\milli\meter}$) in the transverse directions and $n_z
= 256$ points ($\Delta z = \SI{0.547}{\milli\meter}$) in the longitudinal
direction. The space charge computation is performed on a grid with $n_r = 64$
($\Delta r = \SI{0.039}{\milli\meter}$) radial and $n_l = 64$ ($\Delta l =
\SI{2.188}{\milli\meter}$) longitudinal cells. We define
$\bar{\mathbf{x}}_\mathrm{rms}$ and $\bar{\boldsymbol{\epsilon}}_x$ to be
reference solutions computed with refined parameters. More specifically, we
look at two further simulations: The first uses half of the original time step
$\Delta \mathrm{t}$, twice the number of grid points and cells, as well as
double the number of macroparticles $N_\mathrm{p}$. The second one is the
aforementioned reference, which again halves or doubles the parameter values.
We then consider errors defined by
\begin{align*}
    \delta_{x_\mathrm{rms}} &= \max_{1 \leq i \leq N_z}
    \frac{(\mathbf{x}_\mathrm{rms})_i -
    (\bar{\mathbf{x}}_\mathrm{rms})_i}{(\bar{\mathbf{x}}_\mathrm{rms})_i}\\
    \delta_{\epsilon_x} &= \max_{1 \leq i \leq N_z}
    \frac{(\boldsymbol{\epsilon}_x)_i -
    (\bar{\boldsymbol{\epsilon}}_x)_i}{(\bar{\boldsymbol{\epsilon}}_x)_i}
\end{align*}
in the computed statistical quantities with respect to the selected parameters.
The corresponding numerical results are collected in
\autoref{tab:results_astra}. We observe changes below \SI{5}{\percent} between
the reference and the first refinement step, indicating reliable results.

\begin{figure}[b]
    \begin{center}
    \includegraphics[width=0.45\textwidth]{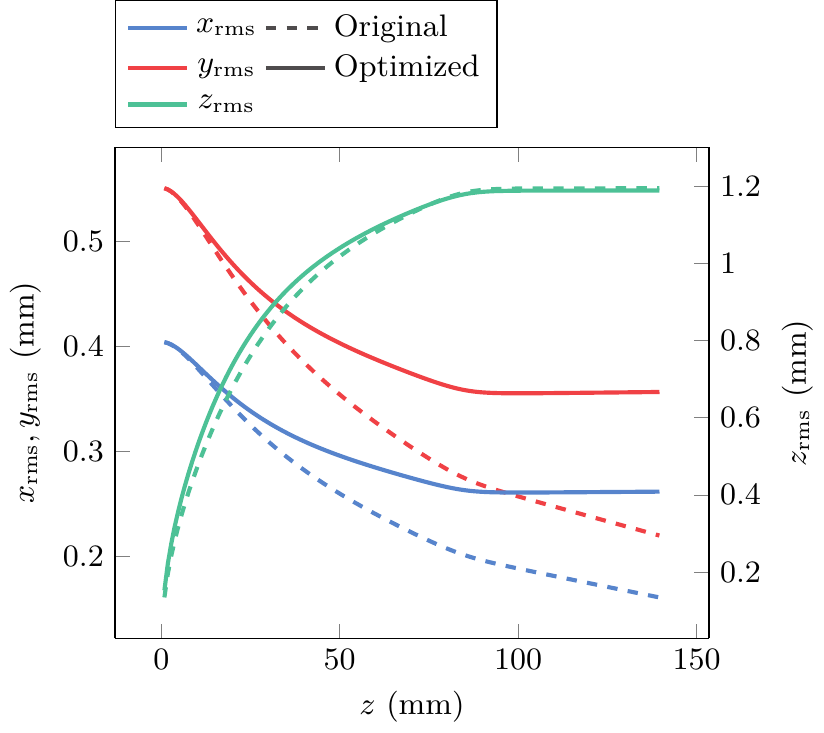}
    \caption{RMS beam widths and RMS beam length for the original and optimized
             geometries as obtained from \textsc{ASTRA} using the reference
             parameters.}
    \label{fig:astra}
    \end{center}
\end{figure}

The computational effort for the tracking simulations has two parts: The
creation of fieldmaps based on the numerical field solution, and the actual
particle tracking procedure. For the initial parameters, the fieldmap
computation took about one hour and the tracking algorithm required around ten
minutes to complete a single run. Thus the fieldmap computation clearly is the
bottleneck in terms of integrating particle simulations into the optimization
procedure. However, it may be possible to obtain sufficiently accurate maps
with significantly less effort and only verify the results afterwards using a
more accurate simulation. The computation times for the refined simulations can
be estimated by a linear scaling, i.e., doubling the number of longitudinal
grid cells or particles roughly doubles the execution time as well. It should
also be noted that the creation of fieldmaps is a highly parallelizable task,
such that considerable speedup could be achieved by a more optimized
implementation.

\begin{table*}[t]
    \begin{ruledtabular}
    \caption{Quantities of interest for the original and optimized geometries.
             All quantities are evaluated at the chamber exit.}
    \begin{tabular}{c|cccccc}
    & $x_\mathrm{rms}$ (\si{\milli\meter}) & $y_\mathrm{rms}$
    (\si{\milli\meter}) & $z_\mathrm{rms}$ (\si{\milli\meter}) & $\epsilon_x$
    (\si{\milli\radian\milli\meter}) & $\epsilon_y$
    (\si{\milli\radian\milli\meter}) & $\epsilon_z$
    (\si{\kilo\electronvolt\milli\meter}) \\
    \hline
    Original & 0.161 & 0.22 & 1.195 & 0.116 & 0.208 & 0.06 \\
    Optimized & 0.258 & 0.36 & 1.189 & 0.108 & 0.194 & 0.06 \\
    \end{tabular}
    \label{tab:results_qoi}
    \end{ruledtabular}
\end{table*}

The numerical solutions for $x_\mathrm{rms}$ and $y_\mathrm{rms}$, interpreted
as functions of $z$, are shown in \autoref{fig:astra}. Also included is the
bunch length $z_\mathrm{rms}$, which may be defined analogously. It can be observed that the optimization slightly worsens
these beam parameters, but not by an intolerable amount. In contrast we even
find a small improvement in the emittance values. Still, this indicates that
future work could achieve yet better results by also considering the shape of
the electrode leading into the photocathode. Our quantities of interest,
evaluated at the chamber exit, are listed in \autoref{tab:results_qoi}. They
match the values measured at operating photo-electron guns
\cite{poltoratska2011, garcia2019}, indicating the reliability of our
simulations.

\begin{figure}[h]
    \begin{center}
    \includegraphics[width=0.45\textwidth]{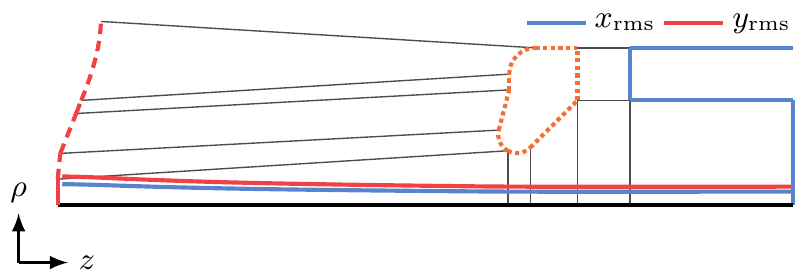}
    \caption{RMS beam widths in relation to part of the geometry, compare
             \autoref{fig:geometry_orig}. The values are scaled up by a factor
             of 10 to indicate that the focussing effect of the electric field
             is sufficiently strong.}
    \label{fig:astra_rms_eps}
    \end{center}
\end{figure}

Lastly, \autoref{fig:astra_rms_eps} shows that the beam fits well inside the
\SI{20}{\milli\meter} aperture of the anode ring, even when considering an
upscaling of the values by a factor of 10 to include a safety margin. In
addition to the values collected in \autoref{tab:results_qoi}, an
RMS energy spread of $\Delta E_\mathrm{rms} = \SI{48.4}{\electronvolt}$ is
observed at the chamber exit for a field strength of
\SI{2.99}{\mega\volt\per\meter} on the photocathode surface at
\SI{-300}{\kilo\volt} bias voltage. Comparing these values to the
\SI{84}{\electronvolt} with \SI{2.5}{\mega\volt\per\meter} at
\SI{-200}{\kilo\volt} bias voltage reported in \cite{friederich2015}, further
supports the validity of our results. It should still be possible to improve on
these values, since the laser shape that was used for the simulations was
unprocessed after emission from the laser diode. Moreover, the procedure
presented in this work focused solely on reducing the maximum electric field
strength on the electrode surface and did not include the full shapes of the
electrode or anode ring in the optimization. Taking their effects on the beam
parameters into account could therefore provide additional improvements. As
mentioned before, the initial momentum of the emitted particles needs to be
included as well, which is expected to increase both RMS emittance and RMS
energy spread. Specific optimization of the emission properties is the focus of
ongoing work.\\

\section{Conclusion and outlook}
A successful IGA-based shape optimization of a DC high-voltage photo-electron
gun was performed. The maximum electric field strength of the optimized
geometry was computed as \SI{9.06}{\mega\volt\per\meter}, which is well below
the field emission threshold of \SI{10}{\mega\volt\per\meter}. This constitutes
a reduction in the maximum field gradient by more than \SI{25}{\percent}
compared to the initial design prior to optimization. Furthermore, the
optimized electrode complies with the weight and volume restrictions, and the
maximum field strength on the cathode surface is determined to be
\SI{2.99}{\mega\volt\per\meter}, which allows for a sufficiently low energy
spread of the electron beam. The procedure was carried out for an electrode
voltage of \SI{-300}{\kilo\volt} and an anode voltage of \SI{1}{\kilo\volt},
with a fixed anode cathode gap of \SI{80}{\milli\meter}. Some beam parameters
of the resulting geometry, namely the RMS beam widths and length, the
normalized transverse and longitudinal RMS beam emittances, and the RMS energy
spread were investigated using the particle tracking software \textsc{ASTRA}.
Preliminary results were found to be in agreement with measurements at
operating guns and exhibited values suitable for the setup at
\mbox{Photo-CATCH}.

The work presented here focused solely on the optimization of the maximum
electric field strength. In the future, the same procedure may be applied to
the entire electrode, and the anode ring also, optimizing both their shapes and
the anode-cathode distance. This includes fully coupling the shape optimization
with a particle tracking software as well, to optimize the emission properties
of the gun and investigate the influence of the electrode geometry on beam
characteristics for ERL-typical bunch charges of \SI{100}{\pico\coulomb} and
above. In this context, it could also prove useful to allow for a direct
evaluation of the spline basis functions within the tracking code, to make full
use of the increased accuracy and smoothness without sacrificing computational
efficiency.

\section*{Acknowledgments}
The authors gratefully acknowledge the fruitful discussions with Simon
Friederich, Yuliya Fritzsche and Steffen Schmid. M. Herbert and J. Enders
acknowledge support by the Deutsche Forschungsgemeinschaft (DFG) -
Projektnummer 264883531 (GRK 2128 ``AccelencE'') and the German BMBF
(05H18RDRB1).

\bibliography{biblio, bibtex_all}

\begin{thebibliography}{60}%
\makeatletter
\providecommand \@ifxundefined [1]{%
 \@ifx{#1\undefined}
}%
\providecommand \@ifnum [1]{%
 \ifnum #1\expandafter \@firstoftwo
 \else \expandafter \@secondoftwo
 \fi
}%
\providecommand \@ifx [1]{%
 \ifx #1\expandafter \@firstoftwo
 \else \expandafter \@secondoftwo
 \fi
}%
\providecommand \natexlab [1]{#1}%
\providecommand \enquote  [1]{``#1''}%
\providecommand \bibnamefont  [1]{#1}%
\providecommand \bibfnamefont [1]{#1}%
\providecommand \citenamefont [1]{#1}%
\providecommand \href@noop [0]{\@secondoftwo}%
\providecommand \href [0]{\begingroup \@sanitize@url \@href}%
\providecommand \@href[1]{\@@startlink{#1}\@@href}%
\providecommand \@@href[1]{\endgroup#1\@@endlink}%
\providecommand \@sanitize@url [0]{\catcode `\\12\catcode `\$12\catcode
  `\&12\catcode `\#12\catcode `\^12\catcode `\_12\catcode `\%12\relax}%
\providecommand \@@startlink[1]{}%
\providecommand \@@endlink[0]{}%
\providecommand \url  [0]{\begingroup\@sanitize@url \@url }%
\providecommand \@url [1]{\endgroup\@href {#1}{\urlprefix }}%
\providecommand \urlprefix  [0]{URL }%
\providecommand \Eprint [0]{\href }%
\providecommand \doibase [0]{https://doi.org/}%
\providecommand \selectlanguage [0]{\@gobble}%
\providecommand \bibinfo  [0]{\@secondoftwo}%
\providecommand \bibfield  [0]{\@secondoftwo}%
\providecommand \translation [1]{[#1]}%
\providecommand \BibitemOpen [0]{}%
\providecommand \bibitemStop [0]{}%
\providecommand \bibitemNoStop [0]{.\EOS\space}%
\providecommand \EOS [0]{\spacefactor3000\relax}%
\providecommand \BibitemShut  [1]{\csname bibitem#1\endcsname}%
\let\auto@bib@innerbib\@empty
\bibitem [{\citenamefont {Sinclair}(2006)}]{sinclair2006}%
  \BibitemOpen
  \bibfield  {author} {\bibinfo {author} {\bibfnamefont {C.~K.}\ \bibnamefont
  {Sinclair}},\ }\bibfield  {title} {\bibinfo {title} {{DC} photoemission
  electron guns as {ERL} sources},\ }\bibfield  {journal} {\bibinfo  {journal}
  {Nuclear Instruments and Methods in Physics Research Section A: Accelerators,
  Spectrometers, Detectors and Associated Equipment}\ }\textbf {\bibinfo
  {volume} {557}},\ \href {https://doi.org/10.1016/j.nima.2005.10.053}
  {10.1016/j.nima.2005.10.053} (\bibinfo {year} {2006})\BibitemShut {NoStop}%
\bibitem [{\citenamefont {Rao}\ \emph {et~al.}(2006)\citenamefont {Rao} \emph
  {et~al.}}]{rao2006}%
  \BibitemOpen
  \bibfield  {author} {\bibinfo {author} {\bibfnamefont {T.}~\bibnamefont
  {Rao}} \emph {et~al.},\ }\bibfield  {title} {\bibinfo {title} {Photocathodes
  for the energy recovery linacs},\ }\bibfield  {journal} {\bibinfo  {journal}
  {Nuclear Instruments and Methods in Physics Research Section A: Accelerators,
  Spectrometers, Detectors and Associated Equipment}\ }\textbf {\bibinfo
  {volume} {557}},\ \href {https://doi.org/10.1016/j.nima.2005.10.112}
  {10.1016/j.nima.2005.10.112} (\bibinfo {year} {2006})\BibitemShut {NoStop}%
\bibitem [{\citenamefont {Maxson}\ \emph {et~al.}(2014)\citenamefont {Maxson}
  \emph {et~al.}}]{maxson2014}%
  \BibitemOpen
  \bibfield  {author} {\bibinfo {author} {\bibfnamefont {J.}~\bibnamefont
  {Maxson}} \emph {et~al.},\ }\bibfield  {title} {\bibinfo {title} {Design,
  conditioning, and performance of a high voltage, high brightness {DC}
  photoelectron gun with variable gap},\ }\bibfield  {journal} {\bibinfo
  {journal} {Review of Scientific Instruments}\ }\textbf {\bibinfo {volume}
  {85}},\ \href {https://doi.org/10.1063/1.4895641} {10.1063/1.4895641}
  (\bibinfo {year} {2014})\BibitemShut {NoStop}%
\bibitem [{\citenamefont {Sinclair}\ \emph {et~al.}(2007)\citenamefont
  {Sinclair} \emph {et~al.}}]{sinclair2007}%
  \BibitemOpen
  \bibfield  {author} {\bibinfo {author} {\bibfnamefont {C.~K.}\ \bibnamefont
  {Sinclair}} \emph {et~al.},\ }\bibfield  {title} {\bibinfo {title}
  {Development of a high average current polarized electron source with long
  cathode operational lifetime},\ }\bibfield  {journal} {\bibinfo  {journal}
  {Physical Review Special Topics Accelerators and Beams}\ }\textbf {\bibinfo
  {volume} {10}},\ \href {https://doi.org/10.1103/PhysRevSTAB.10.023501}
  {10.1103/PhysRevSTAB.10.023501} (\bibinfo {year} {2007})\BibitemShut
  {NoStop}%
\bibitem [{\citenamefont {Pierce}(1940)}]{pierce1940}%
  \BibitemOpen
  \bibfield  {author} {\bibinfo {author} {\bibfnamefont {J.~R.}\ \bibnamefont
  {Pierce}},\ }\bibfield  {title} {\bibinfo {title} {Rectilinear electron flow
  in beams},\ }\bibfield  {journal} {\bibinfo  {journal} {Journal of Applied
  Physics}\ }\textbf {\bibinfo {volume} {11}},\ \href
  {https://doi.org/10.1063/1.1712815} {10.1063/1.1712815} (\bibinfo {year}
  {1940})\BibitemShut {NoStop}%
\bibitem [{\citenamefont {Radley}(1958)}]{radley1958}%
  \BibitemOpen
  \bibfield  {author} {\bibinfo {author} {\bibfnamefont {D.~E.}\ \bibnamefont
  {Radley}},\ }\bibfield  {title} {\bibinfo {title} {The theory of the pierce
  type electron gun},\ }\bibfield  {journal} {\bibinfo  {journal} {Journal of
  Electronics and Control}\ }\textbf {\bibinfo {volume} {4}},\ \href
  {https://doi.org/10.1080/00207215808953831} {10.1080/00207215808953831}
  (\bibinfo {year} {1958})\BibitemShut {NoStop}%
\bibitem [{\citenamefont {Bazarov}\ and\ \citenamefont
  {Sinclair}(2005)}]{bazarov2005}%
  \BibitemOpen
  \bibfield  {author} {\bibinfo {author} {\bibfnamefont {I.~V.}\ \bibnamefont
  {Bazarov}}\ and\ \bibinfo {author} {\bibfnamefont {C.~K.}\ \bibnamefont
  {Sinclair}},\ }\bibfield  {title} {\bibinfo {title} {Multivariate
  optimization of a high brightness dc gun photoinjector},\ }\bibfield
  {journal} {\bibinfo  {journal} {Physical Review Special Topics Accelerators
  and Beams}\ }\textbf {\bibinfo {volume} {8}},\ \href
  {https://doi.org/10.1103/PhysRevSTAB.8.034202} {10.1103/PhysRevSTAB.8.034202}
  (\bibinfo {year} {2005})\BibitemShut {NoStop}%
\bibitem [{\citenamefont {Bazarov}\ \emph {et~al.}(2011)\citenamefont {Bazarov}
  \emph {et~al.}}]{bazarov2011}%
  \BibitemOpen
  \bibfield  {author} {\bibinfo {author} {\bibfnamefont {I.~V.}\ \bibnamefont
  {Bazarov}} \emph {et~al.},\ }\bibfield  {title} {\bibinfo {title} {Comparison
  of dc and superconducting rf photoemission guns for high brightness high
  average current beam production},\ }\bibfield  {journal} {\bibinfo  {journal}
  {Physical Review Special Topics Accelerators and Beams}\ }\textbf {\bibinfo
  {volume} {14}},\ \href {https://doi.org/10.1103/PhysRevSTAB.14.072001}
  {10.1103/PhysRevSTAB.14.072001} (\bibinfo {year} {2011})\BibitemShut
  {NoStop}%
\bibitem [{\citenamefont {Grames}\ \emph {et~al.}(2011)\citenamefont {Grames}
  \emph {et~al.}}]{grames2011}%
  \BibitemOpen
  \bibfield  {author} {\bibinfo {author} {\bibfnamefont {J.}~\bibnamefont
  {Grames}} \emph {et~al.},\ }\bibfield  {title} {\bibinfo {title} {Charge and
  fluence lifetime measurements of a {DC} high voltage gaas photogun at high
  average current},\ }\bibfield  {journal} {\bibinfo  {journal} {Physical
  Review Special Topics Accelerators and Beams}\ }\textbf {\bibinfo {volume}
  {14}},\ \href {https://doi.org/10.1103/PhysRevSTAB.14.043501}
  {10.1103/PhysRevSTAB.14.043501} (\bibinfo {year} {2011})\BibitemShut
  {NoStop}%
\bibitem [{\citenamefont {Breidenbach}\ \emph {et~al.}(1994)\citenamefont
  {Breidenbach} \emph {et~al.}}]{breidenbach1994}%
  \BibitemOpen
  \bibfield  {author} {\bibinfo {author} {\bibfnamefont {M.}~\bibnamefont
  {Breidenbach}} \emph {et~al.},\ }\bibfield  {title} {\bibinfo {title} {An
  inverted-geometry, high voltage polarized electron gun with {UHV} load
  lock},\ }\bibfield  {journal} {\bibinfo  {journal} {Nuclear Instruments and
  Methods in Physics Research Section A: Accelerators, Spectrometers, Detectors
  and Associated Equipment}\ }\textbf {\bibinfo {volume} {350}},\ \href
  {https://doi.org/10.1016/0168-9002(94)91146-0} {10.1016/0168-9002(94)91146-0}
  (\bibinfo {year} {1994})\BibitemShut {NoStop}%
\bibitem [{\citenamefont {Poltoratska}\ \emph {et~al.}(2011)\citenamefont
  {Poltoratska} \emph {et~al.}}]{poltoratska2011}%
  \BibitemOpen
  \bibfield  {author} {\bibinfo {author} {\bibfnamefont {Y.}~\bibnamefont
  {Poltoratska}} \emph {et~al.},\ }\bibfield  {title} {\bibinfo {title} {Status
  and recent developments at the polarized-electron injector of the
  superconducting {D}armstadt electron linear accelerator {S}-{DALINAC}},\
  }\bibfield  {journal} {\bibinfo  {journal} {Journal of Physics: Conference
  Series}\ }\textbf {\bibinfo {volume} {298}},\ \href
  {https://doi.org/10.1088/1742-6596/298/1/012002}
  {10.1088/1742-6596/298/1/012002} (\bibinfo {year} {2011})\BibitemShut
  {NoStop}%
\bibitem [{\citenamefont {Hernandez-Garcia}\ \emph {et~al.}(2019)\citenamefont
  {Hernandez-Garcia} \emph {et~al.}}]{garcia2019}%
  \BibitemOpen
  \bibfield  {author} {\bibinfo {author} {\bibfnamefont {C.}~\bibnamefont
  {Hernandez-Garcia}} \emph {et~al.},\ }\bibfield  {title} {\bibinfo {title}
  {Compact $\ensuremath{-}300\text{ }\mathrm{kV}$ dc inverted insulator
  photogun with biased anode and alkali-antimonide photocathode},\ }\bibfield
  {journal} {\bibinfo  {journal} {Physical Review Accelerators and Beams}\
  }\textbf {\bibinfo {volume} {22}},\ \href
  {https://doi.org/10.1103/PhysRevAccelBeams.22.113401}
  {10.1103/PhysRevAccelBeams.22.113401} (\bibinfo {year} {2019})\BibitemShut
  {NoStop}%
\bibitem [{\citenamefont {Nishimori}\ \emph {et~al.}(2014)\citenamefont
  {Nishimori} \emph {et~al.}}]{nishimori2014}%
  \BibitemOpen
  \bibfield  {author} {\bibinfo {author} {\bibfnamefont {N.}~\bibnamefont
  {Nishimori}} \emph {et~al.},\ }\bibfield  {title} {\bibinfo {title}
  {Experimental investigation of an optimum configuration for a high-voltage
  photoemission gun for operation at $\ensuremath{\ge}500\text{ }\text{
  }\mathrm{kV}$},\ }\bibfield  {journal} {\bibinfo  {journal} {Physical Review
  Special Topics Accelerators and Beams}\ }\textbf {\bibinfo {volume} {17}},\
  \href {https://doi.org/10.1103/PhysRevSTAB.17.053401}
  {10.1103/PhysRevSTAB.17.053401} (\bibinfo {year} {2014})\BibitemShut
  {NoStop}%
\bibitem [{\citenamefont {Stutzman}\ \emph {et~al.}(2003)\citenamefont
  {Stutzman} \emph {et~al.}}]{stutzman2003}%
  \BibitemOpen
  \bibfield  {author} {\bibinfo {author} {\bibfnamefont {M.~L.}\ \bibnamefont
  {Stutzman}} \emph {et~al.},\ }\bibfield  {title} {\bibinfo {title} {A
  comparison of outgassing measurements for three vacuum chamber materials},\
  }in\ \href@noop {} {\emph {\bibinfo {booktitle} {AIP Conference
  Proceedings}}},\ Vol.\ \bibinfo {volume} {671}\ (\bibinfo {organization}
  {American Institute of Physics},\ \bibinfo {year} {2003})\ pp.\ \bibinfo
  {pages} {300--306}\BibitemShut {NoStop}%
\bibitem [{\citenamefont {Adderley}\ \emph {et~al.}(2010)\citenamefont
  {Adderley} \emph {et~al.}}]{adderley2010}%
  \BibitemOpen
  \bibfield  {author} {\bibinfo {author} {\bibfnamefont {P.~A.}\ \bibnamefont
  {Adderley}} \emph {et~al.},\ }\bibfield  {title} {\bibinfo {title}
  {Load-locked {DC} high voltage {GaAs} photogun with an inverted-geometry
  ceramic insulator},\ }\bibfield  {journal} {\bibinfo  {journal} {Physical
  Review Special Topics Accelerators and Beams}\ }\textbf {\bibinfo {volume}
  {13}},\ \href {https://doi.org/10.1103/PhysRevSTAB.13.010101}
  {10.1103/PhysRevSTAB.13.010101} (\bibinfo {year} {2010})\BibitemShut
  {NoStop}%
\bibitem [{\citenamefont {Sinclair}\ \emph {et~al.}(2001)\citenamefont
  {Sinclair} \emph {et~al.}}]{sinclair2001}%
  \BibitemOpen
  \bibfield  {author} {\bibinfo {author} {\bibfnamefont {C.~K.}\ \bibnamefont
  {Sinclair}} \emph {et~al.},\ }\bibfield  {title} {\bibinfo {title} {Dramatic
  reduction of field emission from large area electrodes by plasma-source ion
  implantation},\ }in\ \href
  {https://accelconf.web.cern.ch/p01/PAPERS/ROPB004.PDF} {\emph {\bibinfo
  {booktitle} {Proc. 2001 Particle Accelerator Conference (PAC'01), Chicago,
  IL, USA, June 18-22, 2001}}},\ \bibinfo {series and number} {Particle
  Accelerator Conference}\ (\bibinfo  {publisher} {IEEE Xplore},\ \bibinfo
  {year} {2001})\ pp.\ \bibinfo {pages} {610--612}\BibitemShut {NoStop}%
\bibitem [{\citenamefont {Bastaninejad}\ \emph {et~al.}(2012)\citenamefont
  {Bastaninejad} \emph {et~al.}}]{bastaninejad2012}%
  \BibitemOpen
  \bibfield  {author} {\bibinfo {author} {\bibfnamefont {M.}~\bibnamefont
  {Bastaninejad}} \emph {et~al.},\ }\bibfield  {title} {\bibinfo {title}
  {Evaluation of niobium as candidate electrode material for {DC} high voltage
  photoelectron guns},\ }\bibfield  {journal} {\bibinfo  {journal} {Physical
  Review Special Topics Accelerators and Beams}\ }\textbf {\bibinfo {volume}
  {15}},\ \href {https://doi.org/10.1103/PhysRevSTAB.15.083502}
  {10.1103/PhysRevSTAB.15.083502} (\bibinfo {year} {2012})\BibitemShut
  {NoStop}%
\bibitem [{\citenamefont {Furuta}\ \emph {et~al.}(2005)\citenamefont {Furuta}
  \emph {et~al.}}]{furuta2005}%
  \BibitemOpen
  \bibfield  {author} {\bibinfo {author} {\bibfnamefont {F.}~\bibnamefont
  {Furuta}} \emph {et~al.},\ }\bibfield  {title} {\bibinfo {title} {Reduction
  of field emission dark current for high-field gradient electron gun by using
  a molybdenum cathode and titanium anode},\ }\bibfield  {journal} {\bibinfo
  {journal} {Nuclear Instruments and Methods in Physics Research Section A:
  Accelerators, Spectrometers, Detectors and Associated Equipment}\ }\textbf
  {\bibinfo {volume} {538}},\ \href
  {https://doi.org/10.1016/j.nima.2004.08.131} {10.1016/j.nima.2004.08.131}
  (\bibinfo {year} {2005})\BibitemShut {NoStop}%
\bibitem [{\citenamefont {Kurichiyanil}\ \emph {et~al.}(2019)\citenamefont
  {Kurichiyanil} \emph {et~al.}}]{kurichiyanil2019}%
  \BibitemOpen
  \bibfield  {author} {\bibinfo {author} {\bibfnamefont {N.}~\bibnamefont
  {Kurichiyanil}} \emph {et~al.},\ }\bibfield  {title} {\bibinfo {title} {A
  test system for optimizing quantum efficiency and dark lifetime of {GaAs}
  photocathodes},\ }\href {https://doi.org/10.1088/1748-0221/14/08/p08025}
  {\bibfield  {journal} {\bibinfo  {journal} {Journal of Instrumentation}\
  }\textbf {\bibinfo {volume} {14}}\bibinfo  {number} { (8)}}\BibitemShut
  {NoStop}%
\bibitem [{\citenamefont {Herbert}\ \emph {et~al.}(2018)\citenamefont {Herbert}
  \emph {et~al.}}]{herbert2018}%
  \BibitemOpen
\bibfield  {number} {  }\bibfield  {author} {\bibinfo {author} {\bibfnamefont
  {M.}~\bibnamefont {Herbert}} \emph {et~al.},\ }\bibfield  {title} {\bibinfo
  {title} {{I}nverted geometry photo-electron gun research and development at
  {TU} {D}armstadt},\ }in\ \href
  {https://doi.org/10.18429/JACoW-IPAC2018-THPMK101} {\emph {\bibinfo
  {booktitle} {Proc. 9th International Particle Accelerator Conference
  (IPAC'18), Vancouver, BC, Canada, April 29-May 4, 2018}}},\ \bibinfo {series
  and number} {\bibinfo {series} {International Particle Accelerator
  Conference}\ No.~\bibinfo {number} {9}}\ (\bibinfo  {publisher} {JACoW
  Publishing},\ \bibinfo {year} {2018})\ pp.\ \bibinfo {pages}
  {4545--4547}\BibitemShut {NoStop}%
\bibitem [{\citenamefont {Hinze}\ \emph {et~al.}(2008)\citenamefont {Hinze},
  \citenamefont {Pinnau}, \citenamefont {Ulbrich},\ and\ \citenamefont
  {Ulbrich}}]{Hinze_2008aa}%
  \BibitemOpen
  \bibfield  {author} {\bibinfo {author} {\bibfnamefont {M.}~\bibnamefont
  {Hinze}}, \bibinfo {author} {\bibfnamefont {R.}~\bibnamefont {Pinnau}},
  \bibinfo {author} {\bibfnamefont {M.}~\bibnamefont {Ulbrich}},\ and\ \bibinfo
  {author} {\bibfnamefont {S.}~\bibnamefont {Ulbrich}},\ }\href@noop {}
  {{\selectlanguage {english}\emph {\bibinfo {title} {Optimization with {PDE}
  Constraints}}}},\ Mathematical Modelling: Theory and Applications\ (\bibinfo
  {publisher} {Springer},\ \bibinfo {year} {2008})\BibitemShut {NoStop}%
\bibitem [{\citenamefont {Monk}(2003)}]{Monk_2003aa}%
  \BibitemOpen
  \bibfield  {author} {\bibinfo {author} {\bibfnamefont {P.}~\bibnamefont
  {Monk}},\ }\href@noop {} {{\selectlanguage {english}\emph {\bibinfo {title}
  {Finite Element Methods for {Maxwell}'s Equations}}}}\ (\bibinfo  {publisher}
  {Oxford University Press},\ \bibinfo {year} {2003})\BibitemShut {NoStop}%
\bibitem [{\citenamefont {Di~Barba}(2010)}]{Di-Barba_2010aa}%
  \BibitemOpen
  \bibfield  {author} {\bibinfo {author} {\bibfnamefont {P.}~\bibnamefont
  {Di~Barba}},\ }\href@noop {} {\emph {\bibinfo {title} {Multiobjective Shape
  Design in Electricity and Magnetism}}},\ Lecture Notes in Electrical
  Engineering\ (\bibinfo  {publisher} {Springer},\ \bibinfo {year}
  {2010})\BibitemShut {NoStop}%
\bibitem [{\citenamefont {Lewis}\ \emph {et~al.}(2004)\citenamefont {Lewis},
  \citenamefont {Tran}, \citenamefont {Read},\ and\ \citenamefont
  {Ives}}]{Lewis_2004ac}%
  \BibitemOpen
  \bibfield  {author} {\bibinfo {author} {\bibfnamefont {B.~M.}\ \bibnamefont
  {Lewis}}, \bibinfo {author} {\bibfnamefont {H.~T.}\ \bibnamefont {Tran}},
  \bibinfo {author} {\bibfnamefont {M.~E.}\ \bibnamefont {Read}},\ and\
  \bibinfo {author} {\bibfnamefont {R.}~\bibnamefont {Ives}},\ }\bibfield
  {title} {\bibinfo {title} {Design of an electron gun using computer
  optimization},\ }\href {https://doi.org/10.1109/tps.2004.827572} {\bibfield
  {journal} {\bibinfo  {journal} {{IEEE} Transactions on Plasma Science}\
  }\textbf {\bibinfo {volume} {32}},\ \bibinfo {pages} {1242} (\bibinfo {year}
  {2004})}\BibitemShut {NoStop}%
\bibitem [{\citenamefont {Jiang}\ \emph {et~al.}(2015)\citenamefont {Jiang},
  \citenamefont {Luo}, \citenamefont {Yan},\ and\ \citenamefont
  {Wang}}]{Jiang_2015aa}%
  \BibitemOpen
  \bibfield  {author} {\bibinfo {author} {\bibfnamefont {W.}~\bibnamefont
  {Jiang}}, \bibinfo {author} {\bibfnamefont {Y.}~\bibnamefont {Luo}}, \bibinfo
  {author} {\bibfnamefont {R.}~\bibnamefont {Yan}},\ and\ \bibinfo {author}
  {\bibfnamefont {S.}~\bibnamefont {Wang}},\ }\bibfield  {title} {\bibinfo
  {title} {Genetic algorithm-based shape optimization of modulating anode for
  magnetron injection gun with low velocity spread},\ }\href
  {https://doi.org/10.1109/ted.2015.2443068} {\bibfield  {journal} {\bibinfo
  {journal} {{IEEE} Transactions on Electron Devices}\ }\textbf {\bibinfo
  {volume} {62}},\ \bibinfo {pages} {2657} (\bibinfo {year}
  {2015})}\BibitemShut {NoStop}%
\bibitem [{\citenamefont {Ribton}\ and\ \citenamefont
  {Balachandran}(2015)}]{Ribton_2015aa}%
  \BibitemOpen
  \bibfield  {author} {\bibinfo {author} {\bibfnamefont {C.}~\bibnamefont
  {Ribton}}\ and\ \bibinfo {author} {\bibfnamefont {W.}~\bibnamefont
  {Balachandran}},\ }\bibfield  {title} {\bibinfo {title} {Development of an
  evolutionary algorithm for design of electron guns for material processing},\
  }in\ \href {https://ieeexplore.ieee.org/document/7529317} {\emph {\bibinfo
  {booktitle} {7th International Joint Conference on Computational Intelligence
  ({IJCCI} 2015)}}},\ Vol.~\bibinfo {volume} {1}\ (\bibinfo  {publisher}
  {IEEE},\ \bibinfo {year} {2015})\ pp.\ \bibinfo {pages}
  {138--148}\BibitemShut {NoStop}%
\bibitem [{\citenamefont {Stancari}\ and\ \citenamefont
  {Edelen}(2017)}]{Stancari_2017aa}%
  \BibitemOpen
  \bibfield  {author} {\bibinfo {author} {\bibfnamefont {G.}~\bibnamefont
  {Stancari}}\ and\ \bibinfo {author} {\bibfnamefont {J.}~\bibnamefont
  {Edelen}},\ }\href@noop {} {\bibinfo {title} {Optimization of field at
  cathode with python wrapper and {SPIFFE}}},\ \bibinfo {howpublished}
  {\url{https://cdcvs.fnal.gov/redmine/projects/mcmillan-electron-gun/wiki}}
  (\bibinfo {year} {2017})\BibitemShut {NoStop}%
\bibitem [{\citenamefont {Tulu}\ \emph {et~al.}(2018)\citenamefont {Tulu},
  \citenamefont {van Rienen},\ and\ \citenamefont {Arnold}}]{Tulu_2018aa}%
  \BibitemOpen
  \bibfield  {author} {\bibinfo {author} {\bibfnamefont {E.~T.}\ \bibnamefont
  {Tulu}}, \bibinfo {author} {\bibfnamefont {U.}~\bibnamefont {van Rienen}},\
  and\ \bibinfo {author} {\bibfnamefont {A.}~\bibnamefont {Arnold}},\
  }\bibfield  {title} {\bibinfo {title} {Systematic study of multipactor
  suppression techniques for a superconducting rf gun},\ }\bibfield  {journal}
  {\bibinfo  {journal} {Physical Review Accelerators and Beams}\ }\textbf
  {\bibinfo {volume} {21}},\ \href
  {https://doi.org/10.1103/PhysRevAccelBeams.21.113402}
  {10.1103/PhysRevAccelBeams.21.113402} (\bibinfo {year} {2018})\BibitemShut
  {NoStop}%
\bibitem [{\citenamefont {De~Boor}(1972)}]{deboor1972}%
  \BibitemOpen
  \bibfield  {author} {\bibinfo {author} {\bibfnamefont {C.}~\bibnamefont
  {De~Boor}},\ }\bibfield  {title} {\bibinfo {title} {On calculating with
  {B}-splines},\ }\href@noop {} {\bibfield  {journal} {\bibinfo  {journal}
  {Journal of Approximation theory}\ }\textbf {\bibinfo {volume} {6}} (\bibinfo
  {year} {1972})}\BibitemShut {NoStop}%
\bibitem [{\citenamefont {Piegl}\ and\ \citenamefont
  {Tiller}(1997)}]{Piegl_1997aa}%
  \BibitemOpen
  \bibfield  {author} {\bibinfo {author} {\bibfnamefont {L.}~\bibnamefont
  {Piegl}}\ and\ \bibinfo {author} {\bibfnamefont {W.}~\bibnamefont {Tiller}},\
  }\href@noop {} {{\selectlanguage {english}\emph {\bibinfo {title} {The
  {NURBS} Book}}}},\ \bibinfo {edition} {2nd}\ ed.\ (\bibinfo  {publisher}
  {Springer},\ \bibinfo {year} {1997})\BibitemShut {NoStop}%
\bibitem [{\citenamefont {Boggs}\ \emph {et~al.}(2005)\citenamefont {Boggs},
  \citenamefont {Althsuler}, \citenamefont {Larzelere}, \citenamefont {Walsh},
  \citenamefont {Clay},\ and\ \citenamefont {Hardwick}}]{Boggs_2005aa}%
  \BibitemOpen
  \bibfield  {author} {\bibinfo {author} {\bibfnamefont {P.~T.}\ \bibnamefont
  {Boggs}}, \bibinfo {author} {\bibfnamefont {A.}~\bibnamefont {Althsuler}},
  \bibinfo {author} {\bibfnamefont {A.~R.}\ \bibnamefont {Larzelere}}, \bibinfo
  {author} {\bibfnamefont {E.~J.}\ \bibnamefont {Walsh}}, \bibinfo {author}
  {\bibfnamefont {R.~L.}\ \bibnamefont {Clay}},\ and\ \bibinfo {author}
  {\bibfnamefont {M.~F.}\ \bibnamefont {Hardwick}},\ }\href
  {https://doi.org/10.2172/876325} {{\selectlanguage {english}\emph {\bibinfo
  {title} {{DART} system analysis}}}},\ \bibinfo {type} {Technical Report}\
  (\bibinfo  {institution} {Sandia National Laboratories},\ \bibinfo {year}
  {2005})\BibitemShut {NoStop}%
\bibitem [{\citenamefont {Halbach}(1976)}]{halbach1976}%
  \BibitemOpen
  \bibfield  {author} {\bibinfo {author} {\bibfnamefont {K.}~\bibnamefont
  {Halbach}},\ }\bibfield  {title} {\bibinfo {title} {Superfish? a computer
  program for evaluation of rf cavities with cylindrical symmetry},\
  }\href@noop {} {\bibfield  {journal} {\bibinfo  {journal} {Particle
  Accelerators}\ }\textbf {\bibinfo {volume} {7}},\ \bibinfo {pages} {213}
  (\bibinfo {year} {1976})}\BibitemShut {NoStop}%
\bibitem [{\citenamefont {{Dassault Syst\`{e}mes}}(2019)}]{CST_2019aa}%
  \BibitemOpen
  \bibfield  {author} {\bibinfo {author} {\bibnamefont {{Dassault
  Syst\`{e}mes}}},\ }\href {https://www.cst.com} {{\selectlanguage
  {english}\emph {\bibinfo {title} {{CST} {STUDIO} {SUITE} 2019}}}} (\bibinfo
  {year} {2019})\BibitemShut {NoStop}%
\bibitem [{\citenamefont {Gjonaj}\ \emph {et~al.}(2009)\citenamefont {Gjonaj},
  \citenamefont {Ackermann}, \citenamefont {Lau}, \citenamefont {Weiland},\
  and\ \citenamefont {Dohlus}}]{Gjonaj_2009aa}%
  \BibitemOpen
  \bibfield  {author} {\bibinfo {author} {\bibfnamefont {E.}~\bibnamefont
  {Gjonaj}}, \bibinfo {author} {\bibfnamefont {W.}~\bibnamefont {Ackermann}},
  \bibinfo {author} {\bibfnamefont {T.}~\bibnamefont {Lau}}, \bibinfo {author}
  {\bibfnamefont {T.}~\bibnamefont {Weiland}},\ and\ \bibinfo {author}
  {\bibfnamefont {M.}~\bibnamefont {Dohlus}},\ }\bibfield  {title} {\bibinfo
  {title} {Coupler kicks in the third harmonic module for the {XFEL}},\ }in\
  \href@noop {} {\emph {\bibinfo {booktitle} {Proceedings of {PAC}09}}}\
  (\bibinfo  {publisher} {PAC09},\ \bibinfo {year} {2009})\BibitemShut
  {NoStop}%
\bibitem [{\citenamefont {Cottrell}\ \emph {et~al.}(2009)\citenamefont
  {Cottrell}, \citenamefont {Hughes},\ and\ \citenamefont
  {Bazilevs}}]{Cottrell_2009aa}%
  \BibitemOpen
  \bibfield  {author} {\bibinfo {author} {\bibfnamefont {J.~A.}\ \bibnamefont
  {Cottrell}}, \bibinfo {author} {\bibfnamefont {T.~J.~R.}\ \bibnamefont
  {Hughes}},\ and\ \bibinfo {author} {\bibfnamefont {Y.}~\bibnamefont
  {Bazilevs}},\ }\href {http://books.google.it/books?id=9Q9y0Xtz5E4C} {\emph
  {\bibinfo {title} {Isogeometric Analysis: Toward Integration of {CAD} and
  {FEA}}}}\ (\bibinfo  {publisher} {Wiley},\ \bibinfo {year}
  {2009})\BibitemShut {NoStop}%
\bibitem [{\citenamefont {Nguyen}\ \emph {et~al.}(2014)\citenamefont {Nguyen},
  \citenamefont {Evgrafov}, \citenamefont {Gravesen},\ and\ \citenamefont
  {Lahaye}}]{Nguyen_2014ab}%
  \BibitemOpen
  \bibfield  {author} {\bibinfo {author} {\bibfnamefont {D.~M.}\ \bibnamefont
  {Nguyen}}, \bibinfo {author} {\bibfnamefont {A.}~\bibnamefont {Evgrafov}},
  \bibinfo {author} {\bibfnamefont {J.}~\bibnamefont {Gravesen}},\ and\
  \bibinfo {author} {\bibfnamefont {D.}~\bibnamefont {Lahaye}},\ }\bibfield
  {title} {{\selectlanguage {english}\bibinfo {title} {Iso-geometric shape
  optimization of magnetic density separators}},\ }\href
  {https://doi.org/10.1108/COMPEL-07-2013-0234} {\bibfield  {journal} {\bibinfo
   {journal} {{COMPEL}: The International Journal for Computation and
  Mathematics in Electrical and Electronic Engineering}\ }\textbf {\bibinfo
  {volume} {33}},\ \bibinfo {pages} {1416} (\bibinfo {year}
  {2014})}\BibitemShut {NoStop}%
\bibitem [{\citenamefont {Pels}\ \emph {et~al.}(2015)\citenamefont {Pels},
  \citenamefont {Bontinck}, \citenamefont {Corno}, \citenamefont {De~Gersem},\
  and\ \citenamefont {Sch\"{o}ps}}]{Pels_2015aa}%
  \BibitemOpen
  \bibfield  {author} {\bibinfo {author} {\bibfnamefont {A.}~\bibnamefont
  {Pels}}, \bibinfo {author} {\bibfnamefont {Z.}~\bibnamefont {Bontinck}},
  \bibinfo {author} {\bibfnamefont {J.}~\bibnamefont {Corno}}, \bibinfo
  {author} {\bibfnamefont {H.}~\bibnamefont {De~Gersem}},\ and\ \bibinfo
  {author} {\bibfnamefont {S.}~\bibnamefont {Sch\"{o}ps}},\ }\bibfield  {title}
  {{\selectlanguage {english}\bibinfo {title} {Optimization of a
  {Stern}-{Gerlach} magnet by magnetic field-circuit coupling and isogeometric
  analysis}},\ }\bibfield  {journal} {\bibinfo  {journal} {{IEEE} Transactions
  on Magnetics}\ }\textbf {\bibinfo {volume} {51}},\ \href
  {https://doi.org/10.1109/TMAG.2015.2462806} {10.1109/TMAG.2015.2462806}
  (\bibinfo {year} {2015})\BibitemShut {NoStop}%
\bibitem [{\citenamefont {Merkel}\ \emph {et~al.}(2019)\citenamefont {Merkel},
  \citenamefont {Gangl},\ and\ \citenamefont {Sch\"{o}ps}}]{Merkel_2019aa}%
  \BibitemOpen
  \bibfield  {author} {\bibinfo {author} {\bibfnamefont {M.}~\bibnamefont
  {Merkel}}, \bibinfo {author} {\bibfnamefont {P.}~\bibnamefont {Gangl}},\ and\
  \bibinfo {author} {\bibfnamefont {S.}~\bibnamefont {Sch\"{o}ps}},\
  }\href@noop {} {{\selectlanguage {english}\emph {\bibinfo {title} {Shape
  Optimization of Rotating Electric Machines using Isogeometric Analysis}}}},\
  \bibinfo {type} {Preprint}\ \bibinfo {number} {arxiv:1908.06009}\ (\bibinfo
  {institution} {Cornell University},\ \bibinfo {year} {2019})\BibitemShut
  {NoStop}%
\bibitem [{\citenamefont {Bontinck}\ \emph {et~al.}(2017)\citenamefont
  {Bontinck}, \citenamefont {Corno}, \citenamefont {De~Gersem}, \citenamefont
  {Kurz}, \citenamefont {Pels}, \citenamefont {Sch\"{o}ps}, \citenamefont
  {Wolf}, \citenamefont {de~Falco}, \citenamefont {D\"{o}lz}, \citenamefont
  {V\'{a}zquez},\ and\ \citenamefont {R\"{o}mer}}]{Bontinck_2017ag}%
  \BibitemOpen
  \bibfield  {author} {\bibinfo {author} {\bibfnamefont {Z.}~\bibnamefont
  {Bontinck}}, \bibinfo {author} {\bibfnamefont {J.}~\bibnamefont {Corno}},
  \bibinfo {author} {\bibfnamefont {H.}~\bibnamefont {De~Gersem}}, \bibinfo
  {author} {\bibfnamefont {S.}~\bibnamefont {Kurz}}, \bibinfo {author}
  {\bibfnamefont {A.}~\bibnamefont {Pels}}, \bibinfo {author} {\bibfnamefont
  {S.}~\bibnamefont {Sch\"{o}ps}}, \bibinfo {author} {\bibfnamefont
  {F.}~\bibnamefont {Wolf}}, \bibinfo {author} {\bibfnamefont {C.}~\bibnamefont
  {de~Falco}}, \bibinfo {author} {\bibfnamefont {J.}~\bibnamefont {D\"{o}lz}},
  \bibinfo {author} {\bibfnamefont {R.}~\bibnamefont {V\'{a}zquez}},\ and\
  \bibinfo {author} {\bibfnamefont {U.}~\bibnamefont {R\"{o}mer}},\ }\bibfield
  {title} {{\selectlanguage {english}\bibinfo {title} {Recent advances of
  isogeometric analysis in computational electromagnetics}},\ }\href
  {https://www.compumag.org/wp/newsletter/} {\bibfield  {journal} {\bibinfo
  {journal} {International Compumag Society Newsletter}\ }\textbf {\bibinfo
  {volume} {24}} (\bibinfo {year} {2017})}\BibitemShut {NoStop}%
\bibitem [{\citenamefont {Piegl}\ and\ \citenamefont
  {Tiller}(1987)}]{Piegl_1987aa}%
  \BibitemOpen
  \bibfield  {author} {\bibinfo {author} {\bibfnamefont {L.}~\bibnamefont
  {Piegl}}\ and\ \bibinfo {author} {\bibfnamefont {W.}~\bibnamefont {Tiller}},\
  }\bibfield  {title} {{\selectlanguage {english}\bibinfo {title} {Curve and
  surface constructions using rational {B}-splines}},\ }\href
  {https://doi.org/10.1016/0010-4485(87)90234-X} {\bibfield  {journal}
  {\bibinfo  {journal} {Computer-Aided Design}\ }\textbf {\bibinfo {volume}
  {19}},\ \bibinfo {pages} {485} (\bibinfo {year} {1987})}\BibitemShut
  {NoStop}%
\bibitem [{\citenamefont {Cohen}\ \emph {et~al.}(2001)\citenamefont {Cohen},
  \citenamefont {Riesenfeld},\ and\ \citenamefont {Elber}}]{Cohen_2001aa}%
  \BibitemOpen
  \bibfield  {author} {\bibinfo {author} {\bibfnamefont {E.}~\bibnamefont
  {Cohen}}, \bibinfo {author} {\bibfnamefont {R.~F.}\ \bibnamefont
  {Riesenfeld}},\ and\ \bibinfo {author} {\bibfnamefont {G.}~\bibnamefont
  {Elber}},\ }\href {https://doi.org/10.1201/9781439864203} {\emph {\bibinfo
  {title} {Geometric Modeling with Splines: An Introduction}}}\ (\bibinfo
  {publisher} {CRC Press},\ \bibinfo {year} {2001})\BibitemShut {NoStop}%
\bibitem [{\citenamefont {de~Boor}(2001)}]{de-Boor_2001aa}%
  \BibitemOpen
  \bibfield  {author} {\bibinfo {author} {\bibfnamefont {C.}~\bibnamefont
  {de~Boor}},\ }\href@noop {} {{\selectlanguage {english}\emph {\bibinfo
  {title} {A Practical Guide to Splines}}}},\ \bibinfo {edition} {rev.}\ ed.,\
  \bibinfo {series} {Applied Mathematical Sciences}, Vol.~\bibinfo {volume}
  {27}\ (\bibinfo  {publisher} {Springer},\ \bibinfo {year} {2001})\BibitemShut
  {NoStop}%
\bibitem [{\citenamefont {Spink}\ \emph {et~al.}(2017)\citenamefont {Spink},
  \citenamefont {Claxton}, \citenamefont {de~Falco},\ and\ \citenamefont
  {V\'azquez}}]{nurbs}%
  \BibitemOpen
  \bibfield  {author} {\bibinfo {author} {\bibfnamefont {M.}~\bibnamefont
  {Spink}}, \bibinfo {author} {\bibfnamefont {D.}~\bibnamefont {Claxton}},
  \bibinfo {author} {\bibfnamefont {C.}~\bibnamefont {de~Falco}},\ and\
  \bibinfo {author} {\bibfnamefont {R.}~\bibnamefont {V\'azquez}},\ }\href
  {https://octave.sourceforge.io/nurbs/} {\emph {\bibinfo {title} {{NURBS}
  Package}}} (\bibinfo {year} {2017})\BibitemShut {NoStop}%
\bibitem [{\citenamefont {Jackson}(1998)}]{Jackson_1998aa}%
  \BibitemOpen
  \bibfield  {author} {\bibinfo {author} {\bibfnamefont {J.~D.}\ \bibnamefont
  {Jackson}},\ }\href {https://doi.org/10.1017/CBO9780511760396}
  {{\selectlanguage {english}\emph {\bibinfo {title} {Classical
  Electrodynamics}}}},\ \bibinfo {edition} {3rd}\ ed.\ (\bibinfo  {publisher}
  {Wiley \& Sons},\ \bibinfo {year} {1998})\BibitemShut {NoStop}%
\bibitem [{\citenamefont {Buffa}\ \emph {et~al.}(2015)\citenamefont {Buffa},
  \citenamefont {V\'{a}zquez}, \citenamefont {Sangalli},\ and\ \citenamefont
  {da~Veiga}}]{Buffa_2015aa}%
  \BibitemOpen
  \bibfield  {author} {\bibinfo {author} {\bibfnamefont {A.}~\bibnamefont
  {Buffa}}, \bibinfo {author} {\bibfnamefont {R.~H.}\ \bibnamefont
  {V\'{a}zquez}}, \bibinfo {author} {\bibfnamefont {G.}~\bibnamefont
  {Sangalli}},\ and\ \bibinfo {author} {\bibfnamefont {L.~B.~a.}\ \bibnamefont
  {da~Veiga}},\ }\bibfield  {title} {{\selectlanguage {english}\bibinfo {title}
  {Approximation estimates for isogeometric spaces in multipatch geometries}},\
  }\href {https://doi.org/10.1002/num.21943} {\bibfield  {journal} {\bibinfo
  {journal} {Numerical Methods for Partial Differential Equations}\ }\textbf
  {\bibinfo {volume} {31}},\ \bibinfo {pages} {422} (\bibinfo {year}
  {2015})}\BibitemShut {NoStop}%
\bibitem [{\citenamefont {Simona}\ \emph {et~al.}(2020)\citenamefont {Simona},
  \citenamefont {Bonaventura}, \citenamefont {de~Falco},\ and\ \citenamefont
  {Sch\"{o}ps}}]{Simona_2020aa}%
  \BibitemOpen
  \bibfield  {author} {\bibinfo {author} {\bibfnamefont {A.}~\bibnamefont
  {Simona}}, \bibinfo {author} {\bibfnamefont {L.}~\bibnamefont {Bonaventura}},
  \bibinfo {author} {\bibfnamefont {C.}~\bibnamefont {de~Falco}},\ and\
  \bibinfo {author} {\bibfnamefont {S.}~\bibnamefont {Sch\"{o}ps}},\ }\bibfield
   {title} {{\selectlanguage {english}\bibinfo {title} {{IsoGeometric}
  approximations for electromagnetic problems in axisymmetric domains}},\
  }\href {https://doi.org/10.1016/j.cma.2020.113211} {\bibfield  {journal}
  {\bibinfo  {journal} {Computer Methods in Applied Mechanics and Engineering}\
  }\textbf {\bibinfo {volume} {369}},\ \bibinfo {pages} {113211} (\bibinfo
  {year} {2020})}\BibitemShut {NoStop}%
\bibitem [{\citenamefont {V\'azquez}(2016)}]{geopdes}%
  \BibitemOpen
  \bibfield  {author} {\bibinfo {author} {\bibfnamefont {R.}~\bibnamefont
  {V\'azquez}},\ }\bibfield  {title} {\bibinfo {title} {A new design for the
  implementation of isogeometric analysis in {O}ctave and {M}atlab: {G}eo{PDE}s
  3.0},\ }\bibfield  {journal} {\bibinfo  {journal} {Computers and Mathematics
  with Applications}\ }\textbf {\bibinfo {volume} {72}},\ \href
  {https://doi.org/10.1016/j.camwa.2016.05.010} {10.1016/j.camwa.2016.05.010}
  (\bibinfo {year} {2016})\BibitemShut {NoStop}%
\bibitem [{\citenamefont {{Runarsson}}\ and\ \citenamefont {{Xin
  Yao}}(2005)}]{runarsson2005}%
  \BibitemOpen
  \bibfield  {author} {\bibinfo {author} {\bibfnamefont {T.~P.}\ \bibnamefont
  {{Runarsson}}}\ and\ \bibinfo {author} {\bibnamefont {{Xin Yao}}},\
  }\bibfield  {title} {\bibinfo {title} {Search biases in constrained
  evolutionary optimization},\ }\href@noop {} {\bibfield  {journal} {\bibinfo
  {journal} {IEEE Transactions on Systems, Man, and Cybernetics, Part C
  (Applications and Reviews)}\ }\textbf {\bibinfo {volume} {35}} (\bibinfo
  {year} {2005})}\BibitemShut {NoStop}%
\bibitem [{\citenamefont {Powell}(1994)}]{powell1994}%
  \BibitemOpen
  \bibfield  {author} {\bibinfo {author} {\bibfnamefont {M.~J.~D.}\
  \bibnamefont {Powell}},\ }\bibinfo {title} {A direct search optimization
  method that models the objective and constraint functions by linear
  interpolation},\ in\ \href {https://doi.org/10.1007/978-94-015-8330-5_4}
  {\emph {\bibinfo {booktitle} {Advances in Optimization and Numerical
  Analysis}}}\ (\bibinfo  {publisher} {Springer Netherlands},\ \bibinfo {year}
  {1994})\ pp.\ \bibinfo {pages} {51--67}\BibitemShut {NoStop}%
\bibitem [{\citenamefont {Johnson}(2020)}]{nlopt}%
  \BibitemOpen
  \bibfield  {author} {\bibinfo {author} {\bibfnamefont {S.~G.}\ \bibnamefont
  {Johnson}},\ }\href {http://github.com/stevengj/nlopt} {\emph {\bibinfo
  {title} {The NLopt nonlinear-optimization package Version 2.6.2}}} (\bibinfo
  {year} {2020})\BibitemShut {NoStop}%
\bibitem [{\citenamefont {F\"orster}\ \emph {et~al.}(2021)\citenamefont
  {F\"orster}, \citenamefont {Sch\"ops},\ and\ \citenamefont
  {Simona}}]{egunopt}%
  \BibitemOpen
  \bibfield  {author} {\bibinfo {author} {\bibfnamefont {P.}~\bibnamefont
  {F\"orster}}, \bibinfo {author} {\bibfnamefont {S.}~\bibnamefont
  {Sch\"ops}},\ and\ \bibinfo {author} {\bibfnamefont {A.}~\bibnamefont
  {Simona}},\ }\href@noop {} {\emph {\bibinfo {title} {EgunOpt}}} (\bibinfo
  {year} {2021}),\ \bibinfo {note}
  {\url{https://github.com/temf/EgunOpt}}\BibitemShut {NoStop}%
\bibitem [{\citenamefont {{Siggins}}\ \emph {et~al.}(2001)\citenamefont
  {{Siggins}} \emph {et~al.}}]{siggins2001}%
  \BibitemOpen
  \bibfield  {author} {\bibinfo {author} {\bibfnamefont {T.}~\bibnamefont
  {{Siggins}}} \emph {et~al.},\ }\bibfield  {title} {\bibinfo {title}
  {{Performance of a {DC} {GaAs} photocathode gun for the Jefferson lab
  {FEL}}},\ }\bibfield  {journal} {\bibinfo  {journal} {Nuclear Instruments and
  Methods in Physics Research A}\ }\textbf {\bibinfo {volume} {475}},\ \href
  {https://doi.org/10.1016/S0168-9002(01)01596-0}
  {10.1016/S0168-9002(01)01596-0} (\bibinfo {year} {2001})\BibitemShut
  {NoStop}%
\bibitem [{\citenamefont {Friederich}\ and\ \citenamefont
  {Aulenbacher}(2015)}]{friederich2015}%
  \BibitemOpen
  \bibfield  {author} {\bibinfo {author} {\bibfnamefont {S.}~\bibnamefont
  {Friederich}}\ and\ \bibinfo {author} {\bibfnamefont {K.}~\bibnamefont
  {Aulenbacher}},\ }\bibfield  {title} {\bibinfo {title} {{T}est electron
  source for increased brightness emission by near band gap photoemission},\
  }in\ \href {https://doi.org/10.18429/JACoW-IPAC2015-TUPWA044} {\emph
  {\bibinfo {booktitle} {Proc. 6th International Particle Accelerator
  Conference (IPAC'15), Richmond, VA, USA, May 3-8, 2015}}},\ \bibinfo {series
  and number} {\bibinfo {series} {International Particle Accelerator
  Conference}\ No.~\bibinfo {number} {6}}\ (\bibinfo  {publisher} {JACoW},\
  \bibinfo {year} {2015})\ pp.\ \bibinfo {pages} {1512--1514}\BibitemShut
  {NoStop}%
\bibitem [{\citenamefont {{Hernandez-Garcia}}\ \emph
  {et~al.}(2016)\citenamefont {{Hernandez-Garcia}}, \citenamefont {{Poelker}},\
  and\ \citenamefont {{Hansknecht}}}]{garcia2016}%
  \BibitemOpen
  \bibfield  {author} {\bibinfo {author} {\bibfnamefont {C.}~\bibnamefont
  {{Hernandez-Garcia}}}, \bibinfo {author} {\bibfnamefont {M.}~\bibnamefont
  {{Poelker}}},\ and\ \bibinfo {author} {\bibfnamefont {J.}~\bibnamefont
  {{Hansknecht}}},\ }\bibfield  {title} {\bibinfo {title} {High voltage studies
  of inverted-geometry ceramic insulators for a $\ensuremath350\text{
  }\mathrm{kV}$ {DC} polarized electron gun},\ }\bibfield  {journal} {\bibinfo
  {journal} {IEEE Transactions on Dielectrics and Electrical Insulation}\
  }\textbf {\bibinfo {volume} {23}},\ \href
  {https://doi.org/10.1109/TDEI.2015.005126} {10.1109/TDEI.2015.005126}
  (\bibinfo {year} {2016})\BibitemShut {NoStop}%
\bibitem [{\citenamefont {Palacios-Serrano}\ \emph {et~al.}(2018)\citenamefont
  {Palacios-Serrano} \emph {et~al.}}]{serrano2018}%
  \BibitemOpen
  \bibfield  {author} {\bibinfo {author} {\bibfnamefont {G.}~\bibnamefont
  {Palacios-Serrano}} \emph {et~al.},\ }\bibfield  {title} {\bibinfo {title}
  {{E}lectrostatic design and conditioning of a triple point junction shield
  for a $\ensuremath{-}200\text{ }\mathrm{kV}$ {DC} high voltage photogun},\
  }\bibfield  {journal} {\bibinfo  {journal} {Review of Scientific
  Instruments}\ }\textbf {\bibinfo {volume} {89}},\ \href
  {https://doi.org/10.1063/1.5048700} {10.1063/1.5048700} (\bibinfo {year}
  {2018})\BibitemShut {NoStop}%
\bibitem [{\citenamefont {Floettmann}(2017)}]{astra}%
  \BibitemOpen
  \bibfield  {author} {\bibinfo {author} {\bibfnamefont {K.}~\bibnamefont
  {Floettmann}},\ }\href {https://www.desy.de/~mpyflo/} {\emph {\bibinfo
  {title} {A Space Charge Tracking Algorithm Version 3.2}}},\ \bibinfo
  {organization} {Deutsches Elektronen-Synchrotron DESY} (\bibinfo {year}
  {2017})\BibitemShut {NoStop}%
\bibitem [{\citenamefont {Espig}\ \emph {et~al.}(2014)\citenamefont {Espig},
  \citenamefont {Enders}, \citenamefont {Fritzsche},\ and\ \citenamefont
  {Wagner}}]{espig2014}%
  \BibitemOpen
  \bibfield  {author} {\bibinfo {author} {\bibfnamefont {M.}~\bibnamefont
  {Espig}}, \bibinfo {author} {\bibfnamefont {J.}~\bibnamefont {Enders}},
  \bibinfo {author} {\bibfnamefont {Y.}~\bibnamefont {Fritzsche}},\ and\
  \bibinfo {author} {\bibfnamefont {M.}~\bibnamefont {Wagner}},\ }\bibfield
  {title} {\bibinfo {title} {Investigation of pulsed spin polarized electron
  beams at the {S-DALINAC}},\ }in\ \href {https://doi.org/10.22323/1.182.0059}
  {\emph {\bibinfo {booktitle} {Proceedings of XVth International Workshop on
  Polarized Sources, Targets, and Polarimetry {\textemdash} PoS(PSTP2013)}}},\
  Vol.\ \bibinfo {volume} {182}\ (\bibinfo {year} {2014})\BibitemShut {NoStop}%
\bibitem [{\citenamefont {Bazarov}\ \emph {et~al.}(2008)\citenamefont {Bazarov}
  \emph {et~al.}}]{bazarov2008}%
  \BibitemOpen
  \bibfield  {author} {\bibinfo {author} {\bibfnamefont {I.~V.}\ \bibnamefont
  {Bazarov}} \emph {et~al.},\ }\bibfield  {title} {\bibinfo {title} {Thermal
  emittance and response time measurements of negative electron affinity
  photocathodes},\ }\bibfield  {journal} {\bibinfo  {journal} {Journal of
  Applied Physics}\ }\textbf {\bibinfo {volume} {103}},\ \href
  {https://doi.org/10.1063/1.2838209} {10.1063/1.2838209} (\bibinfo {year}
  {2008})\BibitemShut {NoStop}%
\bibitem [{\citenamefont {Richter}(1996)}]{richter1996}%
  \BibitemOpen
  \bibfield  {author} {\bibinfo {author} {\bibfnamefont {A.}~\bibnamefont
  {Richter}},\ }\bibfield  {title} {\bibinfo {title} {Operational experience at
  the {S-DALINAC}}\ }(\bibinfo  {publisher} {Inst. of Physics Publ.},\ \bibinfo
  {year} {1996})\BibitemShut {NoStop}%
\bibitem [{\citenamefont {Pietralla}(2018)}]{pietralla2018}%
  \BibitemOpen
  \bibfield  {author} {\bibinfo {author} {\bibfnamefont {N.}~\bibnamefont
  {Pietralla}},\ }\bibfield  {title} {\bibinfo {title} {The institute of
  nuclear physics at the {TU} {D}armstadt},\ }\bibfield  {journal} {\bibinfo
  {journal} {Nuclear Physics News}\ }\textbf {\bibinfo {volume} {28}},\ \href
  {https://doi.org/10.1080/10619127.2018.1463013}
  {10.1080/10619127.2018.1463013} (\bibinfo {year} {2018})\BibitemShut
  {NoStop}%
\end{thebibliography}%

\end{document}